\documentclass[sigplan,10pt]{acmart}

\usepackage{array}
\usepackage{acmart-taps}
\usepackage{multirow}
\usepackage{enumitem}
\usepackage{subcaption}
\usepackage{framed}
\usepackage{fontawesome}
\usepackage[nameinlink,noabbrev]{cleveref}

\newcommand{\circled}[1]{\raisebox{-0.05ex}{\ooalign{\hidewidth\raisebox{0.15ex}{\scriptsize #1}\hidewidth\cr\large$\bigcirc$\cr}}}

\usepackage{pifont}
\newcommand{\Letter}{\ding{41}}

\newcommand{\affmark}[1]{\textsuperscript{#1}}

\newcommand{\cfull}{\scalebox{0.8}{\faCircle}}
\newcommand{\chalf}{\scalebox{0.8}{\faAdjust}}
\newcommand{\cnone}{\scalebox{0.8}{\faCircleO}}

\DeclareMathAlphabet{\mathds}{U}{bbold}{m}{n}

\newenvironment{takeawaybox}[3]{%
  \setlength{\FrameRule}{#1}\setlength{\FrameSep}{#2}\setlength{\OuterFrameSep}{#3}%
  \begin{framed}\noindent\ignorespaces}{\end{framed}}

\newenvironment{challengebox}{\begin{takeawaybox}{0.6pt}{4pt}{6pt}}{\end{takeawaybox}}

\newenvironment{lessonbox}{\begin{takeawaybox}{0.6pt}{4pt}{4pt}}{\end{takeawaybox}}

\newif\ifshowrevision
\showrevisionfalse      

\ifshowrevision
  \newcommand{\revise}[1]{{\color{blue}#1}}
  \newenvironment{reviseblock}{\color{blue}}{}
  \newcommand{\revisecolor}{\color{blue}}   
\else
  \newcommand{\revise}[1]{#1}
  \newenvironment{reviseblock}{}{}
  \newcommand{\revisecolor}{}
\fi

\copyrightyear{2026}\acmYear{2026}
\setcopyright{cc}
\setcctype{by-nc-nd}
\acmConference[SIGOPS ATC 2026]{The 2026 ACM SIGOPS Annual Technical Conference}{November 15-18, 2026}{Hong Kong, China}
\acmBooktitle{The 2026 ACM SIGOPS Annual Technical Conference (SIGOPS ATC 2026), November 15-18, 2026, Hong Kong, China}
\acmDOI{10.1145/3828523.3857007}
\acmISBN{979-8-4007-2818-1/2026/11}

\makeatletter
\renewcommand{\@mkauthors}{%
   \gdef\@currentauthors{}
   \global\setbox\mktitle@bx=\vbox{%
       \unvbox\mktitle@bx\par\medskip
       \centering
       {\@authorfont
           Mingxin~Li\affmark{\textdagger\textdaggerdbl*},
           Enge~Song\affmark{\textdaggerdbl*},
           Yueshang~Zuo\affmark{\textdaggerdbl*},
           Xiaodong~Liu\affmark{\textdaggerdbl},
           Rong~Wen\affmark{\S\textdaggerdbl},
           Qiang~Fu\affmark{\P},\\
           Gianni~Antichi\affmark{$\circ$},
           Jian~He\affmark{\textdaggerdbl},
           Jing~Tie\affmark{\textdaggerdbl},
           Zhou~Shao\affmark{\textdaggerdbl},
           Xiaobo~Xue\affmark{\textdaggerdbl},
           Xiong~Xiao\affmark{\textdaggerdbl},
           Luyao~Zhong\affmark{\textdaggerdbl},
           Shaokai~Zhang\affmark{\textdaggerdbl},
           Jiangu~Zhao\affmark{\textdaggerdbl},
           Jianyuan~Lu\affmark{\textdaggerdbl},
           Shize~Zhang\affmark{\textdaggerdbl},
           Xiaoqing~Sun\affmark{\textdaggerdbl},
           Changgang~Zheng\affmark{\textdaggerdbl},
           Zihao~Fan\affmark{\textdaggerdbl},
           Haonan~Li\affmark{\textdaggerdbl},
           Tian~Pan\affmark{\textdaggerdbl},
           Xiaomin~Wu\affmark{\textdaggerdbl},
           Yang~Song\affmark{\textdaggerdbl},
           Xing~Li\affmark{$\star$\textdaggerdbl},
           Biao~Lyu\affmark{\textdaggerdbl},
           Meng~Li\affmark{\textdagger},
           Haipeng~Dai\affmark{\Letter}\affmark{\textdagger},
           Guihai~Chen\affmark{\textdagger},
           Shunmin~Zhu\affmark{\Letter}\affmark{\textdaggerdbl}\par}
       \medskip
       {\@affiliationfont
           \affmark{\textdagger}Nanjing University \enskip
           \affmark{\textdaggerdbl}Alibaba Cloud \enskip
           \affmark{\S}Fudan University \enskip
           \affmark{\P}RMIT University \enskip
           \affmark{$\circ$}Politecnico di Milano \enskip
           \affmark{$\star$}Zhejiang University\par}
       \bigskip}}
\makeatother

\begin{document}

\title{Scaling LLM Agent Tool Access: The MCP Gateway}

\author{Mingxin Li}
\authornote{Mingxin Li, Enge Song, and Yueshang Zuo contributed equally.}
\orcid{0000-0003-3351-8554}
\affiliation{\institution{Nanjing University}\city{Nanjing}\country{China}}
\affiliation{\institution{Alibaba Cloud}\city{Hangzhou}\country{China}}
\email{mingxinli@smail.nju.edu.cn}

\author{Enge Song}
\authornotemark[1]
\orcid{0000-0002-4442-2817}
\affiliation{\institution{Alibaba Cloud}\city{Hangzhou}\country{China}}
\email{enge.seg@alibaba-inc.com}

\author{Yueshang Zuo}
\authornotemark[1]
\orcid{0009-0005-3450-8376}
\affiliation{\institution{Alibaba Cloud}\city{Hangzhou}\country{China}}
\email{zuoyueshang.zys@alibaba-inc.com}

\author{Xiaodong Liu}
\orcid{0009-0003-1791-0676}
\affiliation{\institution{Alibaba Cloud}\city{Hangzhou}\country{China}}
\email{lxd363005@alibaba-inc.com}

\author{Rong Wen}
\orcid{0009-0003-6384-1165}
\affiliation{\institution{Fudan University}\city{Shanghai}\country{China}}
\affiliation{\institution{Alibaba Cloud}\city{Hangzhou}\country{China}}
\email{wenrong.wr@alibaba-inc.com}

\author{Qiang Fu}
\orcid{0000-0002-8067-5571}
\affiliation{\institution{RMIT University}\city{Melbourne}\country{Australia}}
\email{qiang.fu@rmit.edu.au}

\author{Gianni Antichi}
\orcid{0000-0002-6063-4975}
\affiliation{\institution{Politecnico di Milano}\city{Milan}\country{Italy}}
\email{gianni.antichi@polimi.it}

\author{Jian He}
\orcid{0009-0006-8493-4397}
\affiliation{\institution{Alibaba Cloud}\city{Hangzhou}\country{China}}
\email{hejian.hj@alibaba-inc.com}

\author{Jing Tie}
\orcid{0009-0008-4054-114X}
\affiliation{\institution{Alibaba Cloud}\city{Hangzhou}\country{China}}
\email{tiejing.tj@alibaba-inc.com}

\author{Zhou Shao}
\orcid{0009-0004-1762-057X}
\affiliation{\institution{Alibaba Cloud}\city{Hangzhou}\country{China}}
\email{shaozhou@pku.edu.cn}

\author{Xiaobo Xue}
\orcid{0000-0002-5844-6414}
\affiliation{\institution{Alibaba Cloud}\city{Hangzhou}\country{China}}
\email{xxb263051@alibaba-inc.com}

\author{Xiong Xiao}
\orcid{0009-0009-2088-7714}
\affiliation{\institution{Alibaba Cloud}\city{Hangzhou}\country{China}}
\email{xiong.xiaox@alibaba-inc.com}

\author{Luyao Zhong}
\orcid{0009-0006-0475-4047}
\affiliation{\institution{Alibaba Cloud}\city{Hangzhou}\country{China}}
\email{zhongluyao.zly@alibaba-inc.com}

\author{Shaokai Zhang}
\orcid{0009-0008-0904-8031}
\affiliation{\institution{Alibaba Cloud}\city{Hangzhou}\country{China}}
\email{shaokai.zsk@alibaba-inc.com}

\author{Jiangu Zhao}
\orcid{0009-0006-9424-105X}
\affiliation{\institution{Alibaba Cloud}\city{Hangzhou}\country{China}}
\email{jiangu.zjg@alibaba-inc.com}

\author{Jianyuan Lu}
\orcid{0000-0003-2502-8686}
\affiliation{\institution{Alibaba Cloud}\city{Hangzhou}\country{China}}
\email{gaolin.ljy@alibaba-inc.com}

\author{Shize Zhang}
\orcid{0009-0001-4218-7622}
\affiliation{\institution{Alibaba Cloud}\city{Hangzhou}\country{China}}
\email{shize.zsz@alibaba-inc.com}

\author{Xiaoqing Sun}
\orcid{0000-0001-6942-7692}
\affiliation{\institution{Alibaba Cloud}\city{Hangzhou}\country{China}}
\email{sunxiaoqing.sxq@alibaba-inc.com}

\author{Changgang Zheng}
\orcid{0000-0003-1894-722X}
\affiliation{\institution{Alibaba Cloud}\city{Hangzhou}\country{China}}
\email{zhengchanggang.zcg@alibaba-inc.com}

\author{Zihao Fan}
\orcid{0009-0001-6141-0738}
\affiliation{\institution{Alibaba Cloud}\city{Hangzhou}\country{China}}
\email{fanzihao.fzh@alibaba-inc.com}

\author{Haonan Li}
\orcid{0000-0003-1035-5924}
\affiliation{\institution{Alibaba Cloud}\city{Hangzhou}\country{China}}
\email{wenxu.lhn@alibaba-inc.com}

\author{Tian Pan}
\orcid{0000-0001-7718-0669}
\affiliation{\institution{Alibaba Cloud}\city{Hangzhou}\country{China}}
\email{pt282724@alibaba-inc.com}

\author{Xiaomin Wu}
\orcid{0009-0008-5275-226X}
\affiliation{\institution{Alibaba Cloud}\city{Hangzhou}\country{China}}
\email{hengji.wxm@alibaba-inc.com}

\author{Yang Song}
\orcid{0009-0008-2611-757X}
\affiliation{\institution{Alibaba Cloud}\city{Hangzhou}\country{China}}
\email{song288954@alibaba-inc.com}

\author{Xing Li}
\orcid{0009-0009-5379-7512}
\affiliation{\institution{Zhejiang University}\city{Hangzhou}\country{China}}
\affiliation{\institution{Alibaba Cloud}\city{Hangzhou}\country{China}}
\email{lixing.lix@aliyun-inc.com}

\author{Biao Lyu}
\orcid{0000-0002-8096-5528}
\affiliation{\institution{Alibaba Cloud}\city{Hangzhou}\country{China}}
\email{lubiao.lb@alibaba-inc.com}

\author{Meng Li}
\orcid{0000-0001-5764-960X}
\affiliation{\institution{Nanjing University}\city{Nanjing}\country{China}}
\email{meng@nju.edu.cn}

\author{Haipeng Dai}
\authornote{Corresponding authors are Haipeng Dai and Shunmin Zhu.}
\orcid{0000-0003-0545-8187}
\affiliation{\institution{Nanjing University}\city{Nanjing}\country{China}}
\email{haipengdai@nju.edu.cn}

\author{Guihai Chen}
\orcid{0000-0002-6934-1685}
\affiliation{\institution{Nanjing University}\city{Nanjing}\country{China}}
\email{gchen@nju.edu.cn}

\author{Shunmin Zhu}
\authornotemark[2]
\orcid{0000-0001-7756-0952}
\affiliation{\institution{Alibaba Cloud}\city{Hangzhou}\country{China}}
\email{jianghe.zsm@alibaba-inc.com}

\renewcommand{\shortauthors}{Li et al.}

\makeatletter
\let\mcpgw@authornotes\@authornotes
\def\@authornotes{%
   \renewcommand{\thefootnote}{\ifcase\c@footnote\or\ensuremath{*}\or\Letter\fi}%
   \mcpgw@authornotes}
\makeatother

\begin{abstract}
LLM agents increasingly rely on tool calling to complete tasks.
Model Context Protocol (MCP) standardizes tool calling between LLM agents and tool providers.
It reduces integration cost and drives rapid growth in tool scale.
In production deployments, tool access becomes difficult to scale. 
On the tool provider side, legacy services are not directly callable through MCP; the rapid protocol development creates ongoing compatibility cost.
On the agent side, the number of mounted tools is limited by the LLM context window and inference overhead; mounting a large tool set increases token usage and inference latency and can reduce task success rate.
Moreover, for stateful MCP backends with multiple replicas, preserving session affinity increases client-side complexity.

We present a cloud-scale MCP Gateway system.
It breaks the direct-connect model on the data plane and offloads legacy service integration, consolidating incompatible MCP variants, access control, tool recommendation, and session-aware routing at the gateway.
\revise{The conversion module supports 315 OpenAPI services in the cloud and integrates MCP transport variants.}
\revise{Session-aware routing decouples the scalability of the gateway layer and the MCP service layer.}
\revise{The hybrid retrieval engine sustains $\sim$98\% Top-15 recall; it scales agent tool access to 3{,}000+ tools with improved tool selection accuracy.}
\revise{We also share the lessons learned from deploying the gateway system in production.}

\end{abstract}



\maketitle

\section{Introduction}

\aptLtoX{\begin{figure}[tbp]
\centering
\includegraphics[width=0.255\linewidth]{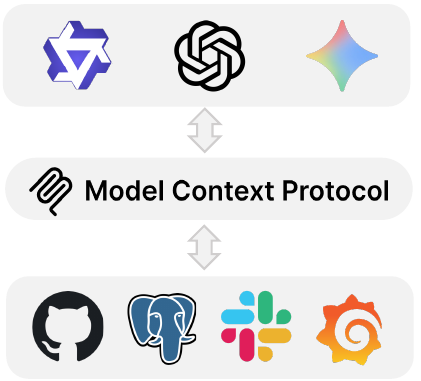}
\caption{MCP}
\label{fig:mcp-before-and-after}
\end{figure}

\begin{figure}[tbp]
\centering
\includegraphics[width=0.51\linewidth]{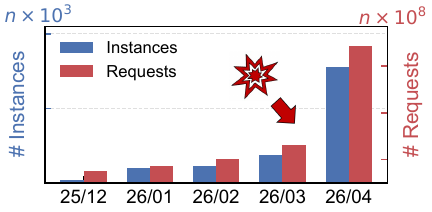}
\caption{Traffic in one cloud region}
\label{fig:mcp_traffic_growth}
\end{figure}}{\begin{figure}[tbp]
  \centering
  \begin{minipage}[b]{0.3\linewidth}
    \centering
    \includegraphics[width=0.85\linewidth]{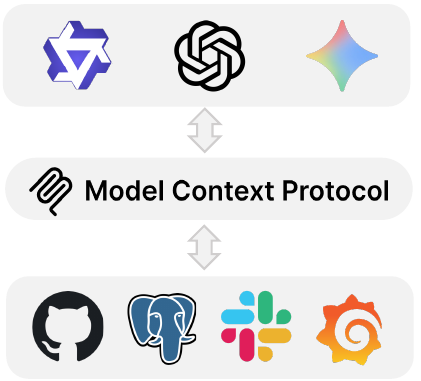}
    \caption{MCP}
    \label{fig:mcp-before-and-after}
  \end{minipage}
  \begin{minipage}[b]{0.6\linewidth}
    \centering
    \includegraphics[width=0.85\linewidth]{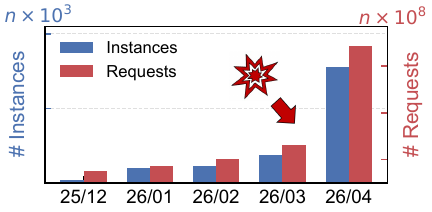}
    \caption{Traffic in one cloud region}
    \label{fig:mcp_traffic_growth}
  \end{minipage}
\end{figure}}


Large Language Model (LLM) applications are evolving from traditional chatbots to tool-using agents.
Tool calling allows an LLM to access external systems and complete complex tasks, such as querying databases or managing cloud network resources~\cite{wu2024netllm,wang2025intent_driven,lin2024parrot,wang2024netassistant,wang2025towards,qiang2024agent_om,fan2025autoprep,li2024llmfordata, lin2025llm}.
Without a unified tool interface, each LLM agent had to be adapted to individual tools. This motivates standardizing the tool-invocation interface, resulting in the Model Context Protocol (MCP)~\cite{mcp_spec_2024_11_05}.
MCP standardizes both how an agent invokes tools and how tool providers expose tools to agents (Figure~\ref{fig:mcp-before-and-after}).
This lowers the engineering overhead for tool providers and enables rapid growth in the number of tools available to agents (reaching $O(10k)$ MCP servers within 13 months~\cite{mcp_so_2025_01_06, mcp_market_2025_01_06}). 
Also, MCP traffic in the cloud is growing rapidly (Figure~\ref{fig:mcp_traffic_growth}).

\begin{reviseblock}
    
\end{reviseblock}
However, building an agent that can access massive tools with reliable tool invocation faces several challenges.
\begin{reviseblock}
(1) The number of public APIs has reached $10^6$~\cite{serbout2024APIstic}, yet these existing web era APIs are not directly callable by agents through MCP.
(2) Tool developers must handle work orthogonal to tool functionality, such as MCP compatibility issues and the integration of authentication.
(3) A single agent can only mount a limited number of tools, insufficient for diverse tasks at scale. 
Mounting too many tools may exceed the context window of the LLM, increasing token cost and latency while reducing task success rate.
(4) MCP assumes a client--server direct-connect model.
For availability and load balancing, servers are often deployed with multiple replicas.
The Host must ensure \emph{session-aware routing} for stateful servers (i.e., all requests in the same logical session are forwarded to the replica that maintains the context).
\end{reviseblock}

To address these issues, we introduce an MCP Gateway, breaking the direct-connect model.
While MCP runs over HTTP, its key routing information resides in the HTTP body rather than headers, preventing existing web-oriented application load balancers (e.g., L7 load balancers and API Gateways) \cite{envoy_latest_docs, nginx_latest_docs, haproxy_latest_docs, apisix_latest_docs} from being directly reused without non-trivial customization.
We implement the following capabilities in the MCP Gateway.
(1) \emph{Protocol Adaptation.} We build a translation module that converts MCP tool calls into calls to existing API servers (enabling access to legacy APIs) and bridges incompatible MCP variants to reduce adaptation cost.
(2) \emph{Function Offloading.} To support fine-grained tool access control at scale, we implement authentication modules that enable identity-to-tool mappings. 
(3) \emph{Tool Recommendation.} We design retrieval methods for MCP.
Beyond naive matching, we exploit the fact that MCP tool schemas often explicitly encode functionality and propose a hybrid matching method combining semantic and lexical matching.
Compared to LLM-based recommendation, this approach is deterministic, efficient, and scalable, making it suitable for deployment at the gateway.
To address the drop in recommendation accuracy when generic matching fails to capture domain-specific knowledge, we further design a query-rewrite method and a knowledge cache based on domain knowledge extraction.
(4) \emph{Session-aware Routing.} The gateway offloads request distribution from the host.
To route requests to backend replicas that hold the corresponding session context, the gateway maintains a session-to-server mapping.
Since cloud gateways are deployed with multiple instances, the gateway synchronizes the session-to-server mapping via a centralized store. 
Our key contributions are:
\begin{itemize}[leftmargin=*]
    \item \textbf{We present the deployed cloud-scale MCP Gateway system for massive tool access.}
    Protocol adaptation enables direct access to legacy APIs with low conversion overhead (P50 $\leq$ 143\,$\mu$s at 100 fields). The tool recommendation module achieves $\sim$98\% recall for Top-15 tools when users mount $O(1k)$ tools, with retrieval latency below 250\,ms. \revise{This reduces time cost and improves agent tool selection accuracy with Top-15 tool recommendations.}
    \item \textbf{We address stateful request routing in multi-gateway deployments.}
    A centralized store-based session synchronization is proposed to support session-aware routing. 
    \revise{We further introduce a Pub/Sub mechanism to synchronize responses across gateway instances to preserve long-lived connections for stateful backends.}
    We show the overhead and explore the trade-offs of session management.
    \item \textbf{We share production deployment lessons.} 
    (1) Maintaining long-lived gateway--backend connections reduces mean latency by up to 51.7\%. (2) Leveraging tool dependencies and locality, knowledge caching returns complete prerequisite tool chains, \revise{and LRU semantic caching achieves up to a 2.29$\times$ speedup in average recommendation latency per workload.}
    \revise{(3) We further discuss an encoding-based session synchronization for cross-gateway coordination, and (4) propose a rule-extraction method, improving tool recommendation Top-15 recall to 99.7\%.}
\end{itemize}



\section{Background and Motivation}\label{sec:background}

\subsection{Tool-Using LLM Agents}

LLM-powered applications are evolving from conversational chatbots to general-purpose autonomous agents~\cite{luo2025autellix,gim2025pie,zheng2024sglang,lin2024parrot}. 
Unlike early chatbots that focused on text generation without external interaction, LLM agents can plan, use tools, and accomplish complex tasks. 
Using reasoning-and-acting paradigms, e.g., ReAct, CoT, and ToT, agents decompose user intents into multi-step tasks~\cite{yao2023react,wei2022cot,yao2023tot} and interact with external systems, e.g., querying databases, managing cloud resources, and triggering workflows~\cite{wu2024netllm,wang2025intent_driven,lin2024parrot,wang2024netassistant,wang2025towards,qiang2024agent_om,fan2025autoprep,li2024llmfordata}. 
The LLM acts as a controller that selects tools, emits structured arguments, and triggers execution (i.e., tool calls)~\cite{patil2024gorilla, schick2023toolformer, woflein2025llm_agents}. 
As a result, LLM agents require robust tool infrastructure to mediate agent–tool interactions~\cite{chan2025infra_agent}. 
The absence of widely adopted protocols remains a major bottleneck to interoperating with external tools and scaling agent ecosystems, motivating recent standardization efforts~\cite{yang2025a_survey_agent_protocols}. 
The MCP~\cite{docs2025mcp} addresses this challenge by defining a standard interface between agents (\emph{MCP Hosts}) and external resources (\emph{MCP Servers}), facilitating tool integration.

\subsection{The Rise of Model Context Protocol}\label{sec:the_rise_of_mcp}

\noindent\textbf{Interoperability via interface standardization.}
Before the MCP, LLM applications and agent frameworks~\cite{langchain,gao2024agentscope,gao2025agentscope} typically integrated external tools through bespoke adapters, resulting in a fragmented ecosystem~\cite{hou2025model}.
\revise{Heterogeneous tool interfaces and provider-specific calling schemas forced point-to-point integration, so connecting $M$ LLM applications to $N$ tools conceptually approached an $O(M \times N)$ engineering burden.
MCP reduces this complexity to $O(M + N)$ by standardizing tool infrastructures: tools are exposed through a uniform, provider-agnostic interface, enabling LLM applications to access diverse tools without per-tool or per-provider rewrites~\cite{docs2025mcp}.}
MCP standardizes interoperability rather than meaning: it does not prescribe tool semantics, backend implementations, or the host’s planning and orchestration logic.



\begin{figure}[!t]
  \centering
  \includegraphics[width=0.8\linewidth]{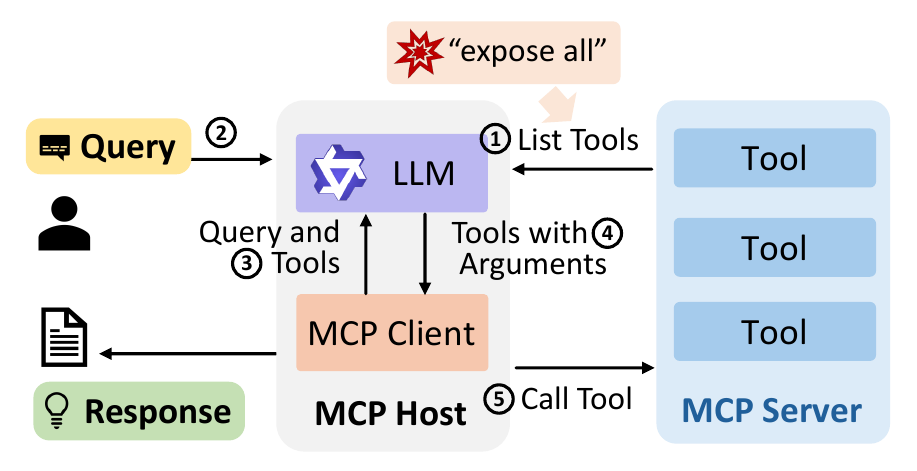}
  
  \caption{
    A concise MCP interaction workflow
  }
  \Description{}
  \label{fig:mcp_workflow_concise}
\end{figure}


\noindent\textbf{MCP interaction workflow.}
An MCP host (e.g., an LLM application or agent) maintains a dedicated MCP client for each connected MCP server.
As shown in \Cref{fig:mcp_workflow_concise}, the workflow starts with connection establishment and initialization.
The host invokes \textit{List Tools} to fetch tool definitions from one or more servers (\circled{1}).
Given a user query (\circled{2}), the host combines the query with these definitions to form a composite prompt (\circled{3}), then invokes the LLM to select tools and produce structured arguments (\circled{4}).
The host parses the arguments and executes the selected tools via \textit{Call Tool} (\circled{5}).
It repeats this select--execute loop until the LLM judges the task completed and returns the final response.

\noindent\textbf{MCP protocol layers.}
MCP defines message semantics at two layers:
(1) \textbf{Data Layer}, representing MCP primitives (e.g., \textit{tools}, \textit{resources}, \textit{prompts}, and \textit{notifications}) and lifecycle management as JSON-RPC messages; and (2) \textbf{Transport Layer}, specifying how these JSON-RPC messages are exchanged between clients and servers.
The transport layer currently supports two options: HTTP+SSE and Streamable HTTP.
\emph{HTTP+SSE} is inherently stateful: the client establishes a long-lived SSE channel for server-to-client delivery, sends client-to-server JSON-RPC via HTTP POST, and includes a server-assigned Session-ID on each request; 
responses are streamed over the single, persistent SSE connection.
\emph{Streamable HTTP} uses a single MCP endpoint, and supports both stateless and stateful modes: 
in stateless mode, each POST is independent and no session is maintained; 
in stateful mode, the server assigns a Session-ID to maintain a logical session across requests without requiring a long-lived connection. 
These transport variants and session semantics directly shape the deployment challenges we characterize next (\S\ref{sec:challenges_of_deploying_mcp}).

\subsection{Challenges of Deploying MCP}
\label{sec:challenges_of_deploying_mcp}

\noindent\textbf{Legacy infrastructures.}
Legacy cloud platforms are primarily built around OpenAPI/Swagger endpoints, engineered for developers and conventional SDKs rather than native tool use by autonomous agents.
This creates a structural mismatch in maturity and supply.
OpenAPI-based infrastructure has been evolving for more than a decade since the first specification release~\cite{swagger_first_reslease}, whereas MCP has existed for only about two years since its public release~\cite{mcp_first_reslease}.
The MCP community registry currently contains on the order of $10^4$ MCP servers~\cite{mcp_market_2025_01_06, mcp_so_2025_01_06}.
This is negligible compared with the \mbox{OpenAPI} ecosystem: as of January 2024, APIstic reports over $10^6$ ($1{,}275{,}568$) valid OpenAPI/Swagger specifications mined from public sources~\cite{serbout2024APIstic}.
Alibaba Cloud exposes on the order of $10^4$--$10^5$ OpenAPI operations~\cite{wang2023gdoc,wang2023improving,alibab_cloud_api_docs}, suggesting that the OpenAPI surface is orders of magnitude larger than the current MCP.
A practical solution must bridge legacy APIs into tool use in a cost-effective and non-intrusive manner.

\begin{challengebox}
\textit{\textbf{Challenge 1:} Large-scale legacy OpenAPI infrastructures are not MCP-native, making most production APIs inaccessible to agent tool use.}
\end{challengebox}

\noindent\textbf{Incompatible protocol versions.}
MCP is evolving rapidly, with four iterations released over a 13-month period, and MCP deployments already face incompatibilities along two fast-moving axes.
First, transport divergence. 
The ecosystem currently spans both the legacy HTTP+SSE transport and the newer Streamable HTTP transport (\S\ref{sec:the_rise_of_mcp}). Measurements from a commercial MCP hosting platform with 184 deployed servers~\cite{mcp_bailian_2025_01_17} show that only a small fraction of servers support both transports, while the rest support only one. 
Client support also lags behind: in the public market, over 50\% of MCP clients support only HTTP+SSE and lack effective support for Streamable HTTP~\cite{guo2025measurement}.
Second, authentication divergence. 
Early MCP specifications did not define a standard authentication mechanism~\cite{mcp_spec_2024_11_05}, leading servers to adopt heterogeneous, vendor-specific security methods.
As authentication requirements evolve, deployments must absorb this heterogeneous authentication ecosystem to preserve compatibility, rather than requiring server developers to repeatedly re-implement authentication logic.

\begin{challengebox}
\textit{\textbf{Challenge 2:} MCP evolution causes transport and authentication incompatibilities across versions, fragmenting deployments and increasing compatibility overhead.}
\end{challengebox}



\begin{table}[tbp]
    \centering
    \footnotesize
    \setlength{\tabcolsep}{4pt}
    \renewcommand{\arraystretch}{0.68}
        \caption{Context window of frontier foundation models, along with the maximum number of tools that can be mounted in model context window\protect\footnotemark.}
    \label{tab:mainstream-context-window}
    \begin{tabular}{@{}p{3.2cm} >{\centering\arraybackslash}p{2.20cm} >{\centering\arraybackslash}p{2.05cm}@{}}
      \toprule
      \textbf{Model} &
      \textbf{Context Window} &
      \textbf{Max \#Tools} \\
      \midrule
      Gemini-3-Flash~\cite{gemini_api_docs}                & \textbf{1M}   & 774 \\
      Qwen-Plus~\cite{qwen_plus_2025_12_01_docs}           & \textbf{1M}   & 627 \\
      GPT-5.2~\cite{gpt52_2025_12_01_docs}                 & \textbf{400k} & 270 \\
      Qwen3-MAX~\cite{qwen3_max_2025_09_23_docs}           & \textbf{256k} & 198 \\
      Claude-Opus-4.5~\cite{anthropic_claude_opus}\cite{anthropic_claude_opus45_news}  & \textbf{200k} & 125 \\
      DeepSeek-V3.2~\cite{deepseek_2025_12_01_docs}        & \textbf{128k} & 98  \\
      \bottomrule
    \end{tabular}
\end{table}
\footnotetext{Measured in our experiments (January 2026) using a tool set derived from Alibaba Cloud VPC, ECS, and other services.}




\begin{figure*}[tbp]
    \centering
    \begin{subfigure}[b]{0.237\textwidth}
      \centering
      \includegraphics[width=\linewidth]{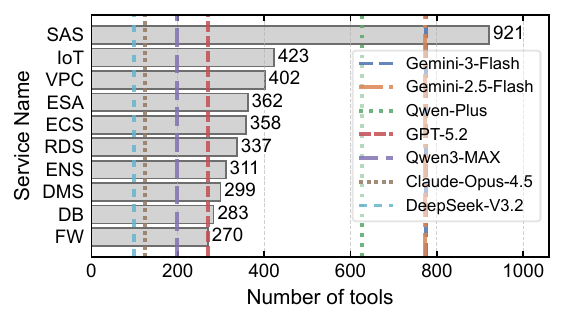}
      \caption{}
      \Description{}
      \label{fig:api-count}
    \end{subfigure}
    \hfill
    \begin{subfigure}[b]{0.237\textwidth}
      \centering
      \includegraphics[width=\linewidth]{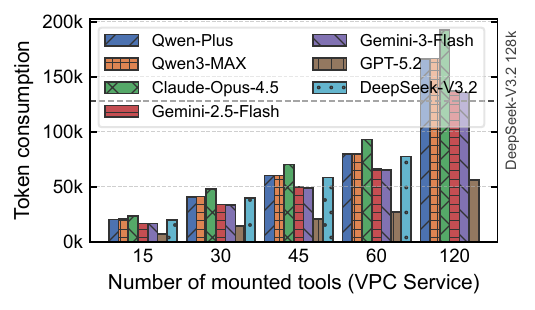}
      \caption{}
      \Description{}
      \label{fig:token-cost}
    \end{subfigure}
    \hfill
    \begin{subfigure}[b]{0.237\textwidth}
      \centering
      \includegraphics[width=\linewidth]{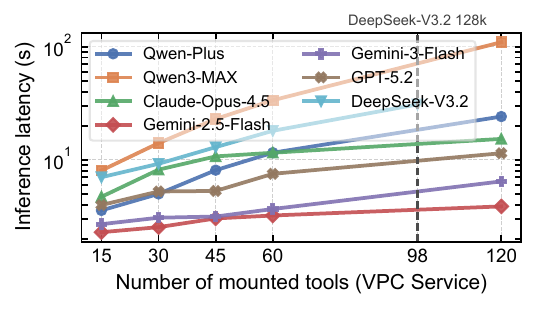}
      \caption{}
      \Description{}
      \label{fig:latency}
    \end{subfigure}
    \hfill
    \begin{subfigure}[b]{0.237\textwidth}
      \centering
      \includegraphics[width=\linewidth]{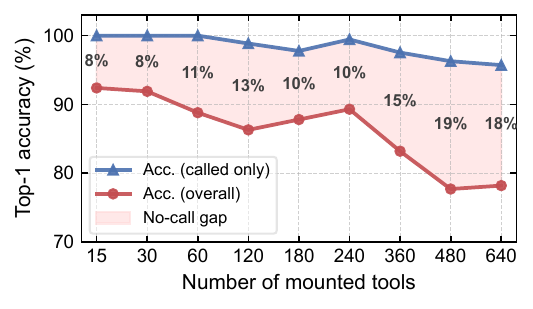}
      \caption{}
      \Description{}
      \label{fig:llm_accuracy}
    \end{subfigure}

    \caption{The measured impact of tool scale.
      \textbf{(a)} \# tools may exceed the capacity of models.
      \textbf{(b)} Inlining schemas incurs  context overhead.
      \textbf{(c)} The scale of the tools leads to huge inference latency.
      \textbf{(d)} Tool-selection accuracy degrades.
    }
    \Description{}
    \label{fig:scale-impacts}
    
  \end{figure*}




\begin{table}[tbp]
    \centering
    \footnotesize
    \setlength{\tabcolsep}{4pt}
    \renewcommand{\arraystretch}{0.68}
    \caption{
        For the VPC agent with $\sim$60 tools, a duplicating VPC task requires $\sim$10 tool-selection rounds.
        In round 1, the KV cache misses, while in rounds 2--10 it hits.
    }\Description{}
    \label{tab:token-cost-60-tools}
    \begin{tabular}{@{}l c c@{}}
      \toprule
      \textbf{Model} & \textbf{Token Usage} & \textbf{Estimated Price} \\
      \midrule
      Claude-Opus-4.5~\cite{anthropic_claude_opus}\cite{anthropic_claude_opus45_news}         & 92{,}924 $\times$ 10 & \$0.999 \\
      Qwen3-MAX~\cite{qwen3_max_2025_09_23_docs}           & 80{,}085 $\times$ 10 & \$0.685 \\
      GPT-5.2~\cite{gpt52_2025_12_01_docs}                 & 27{,}100 $\times$ 10 & \$0.090 \\
      Gemini-3-Flash~\cite{gemini_api_docs}                & 65{,}016 $\times$ 10 & \$0.062 \\
      DeepSeek-V3.2~\cite{deepseek_2025_12_01_docs}        & 77{,}564 $\times$ 10 & \$0.041 \\
      Qwen-Plus~\cite{qwen_plus_2025_12_01_docs}           & 79{,}947 $\times$ 10 & \$0.090 \\
      \bottomrule
    \end{tabular}
\end{table}


\noindent\textbf{The expose-all tool approach.}
MCP hosts (e.g., LLM applications or agents) typically inline all tool schemas into the prompt (\S\ref{sec:the_rise_of_mcp}) and rely on the LLM to select tools and generate arguments.
Although state-of-the-art models support increasingly large context windows, context remains a constrained resource.
Table~\ref{tab:mainstream-context-window} shows that even models with 1M-token context windows effectively saturate at roughly 600--800 tools. 
Because the context window must also accommodate history, this expose-all approach fails to scale to hyperscale tool sets for four reasons.
\textbf{(1) Limited tool capacity.} 
Modern cloud platforms expose a broad range of services and thousands of distinct operations. Even a single service category can include hundreds of tools (Figure~\ref{fig:api-count}), making it difficult for models to accommodate tools from multiple categories. For example, an agent supporting only Security Center Service already requires 921 tools, exceeding the practical capacity of all frontier models.
\textbf{(2) Excessive schema token usage.} 
Tool schemas consume a disproportionate share of the context budget. Inlining just 120 VPC tools requires over 166k tokens for Qwen models, or more than 56k tokens even with efficient tokenization (Figure~\ref{fig:token-cost}). This verbosity increases inference cost (Table~\ref{tab:token-cost-60-tools}) and crowds out agent tasks. 
\textbf{(3) Prohibitive inference latency.} Prompt size grows with the tools, slowing tool selection and argument generation. Figure~\ref{fig:latency} shows that for Qwen3-MAX, mounting 120 tools already pushes latency (prompting + tool selection + argument generation) beyond 100s, while 180 tools exceed 200s. Such delays break interactivity.
\textbf{(4) Degraded accuracy of tool use.} As tool schemas occupy an increasing fraction of the context, models become less reliable at selecting correct tools and generating valid arguments, as shown in Figure~\ref{fig:llm_accuracy}. Specifically, the model increasingly selects inappropriate tools or produces malformed invocations, and, more notably, refuses to call any tool (no-call rate: 8.0\%$\to$19.0\%), degrading agent performance. These failures are analogous to the well-known ``lost-in-the-middle'' phenomenon~\cite{liu2024lost, blankenstein2025biasbusters}.

\begin{challengebox}
\textit{\textbf{Challenge 3:} The expose-all tool approach does not scale: large tool sets overwhelm the context window and incur prohibitive token cost, latency, and accuracy degradation.}
\end{challengebox}

\looseness=-1
\noindent\textbf{Distributing MCP requests across replicas is non-trivial.}
To achieve high availability and throughput with a massive number of tool calls, MCP services typically run multiple MCP server replicas and must distribute requests across them.
In stateful deployments, MCP servers maintain session state across requests, so routing must ensure that all requests in the same session go to the replica holding the context (i.e., \emph{session affinity}).
This requirement makes conventional per-request load-balancing (e.g., round-robin) insufficient.
As a result, hosts/clients often need to implement their own session-aware load-balancing and health-check/failover modules, adding development and operational overhead. 

\begin{challengebox}
\textit{\textbf{Challenge 4:} Stateful MCP sessions make load balancing and replication difficult, requiring session affinity and failure-aware routing to preserve per-session context.}
\end{challengebox}

\subsection{Design Goals}\label{sec:design-goals}

\noindent\textbf{Enabling legacy API access via MCP.}
To address Challenge~1, the system should expose large-scale API operations as standard MCP tools without backend refactoring.

\noindent\textbf{MCP server adaptation overhead reduction.}
To address Challenge~2, the system must act as a compatibility layer across heterogeneous MCP transports and authentication schemes, without modifications to either clients or backends.

\noindent\textbf{LLM context-efficient tool use.}
To address Challenge~3, the system should scale tool access to large catalogs without exhausting the LLM context.
The design goal is bounded context overhead, stable latency and cost, and robust tool selection/argument generation accuracy as the tool set grows.

\noindent\textbf{Request distribution with session affinity.}
To address Challenge~4, the system should distribute requests across MCP replicas while preserving session affinity, without requiring clients to implement session-aware logic.



\section{System Design}\label{sec:system_design}

\noindent\textbf{The MCP needs a gateway system.} In the web-service era, Layer-7 gateways became a foundational building block for reliable client--server communication by centralizing connectivity management, load balancing, and governance.
A similar shared point is needed in the agent-service era, where agents interact with heterogeneous tool infrastructures (e.g., API servers and MCP servers).
Motivated by these needs and informed by our experience in networked services and Layer-7 load balancing as a major cloud provider, we design and deploy the \textbf{MCP Gateway System}.
The gateway breaks the original direct-connect model by interposing a shared control point between MCP clients and tool servers.

\begin{figure}[tbp]
    \centering
    \includegraphics[width=0.95\linewidth]{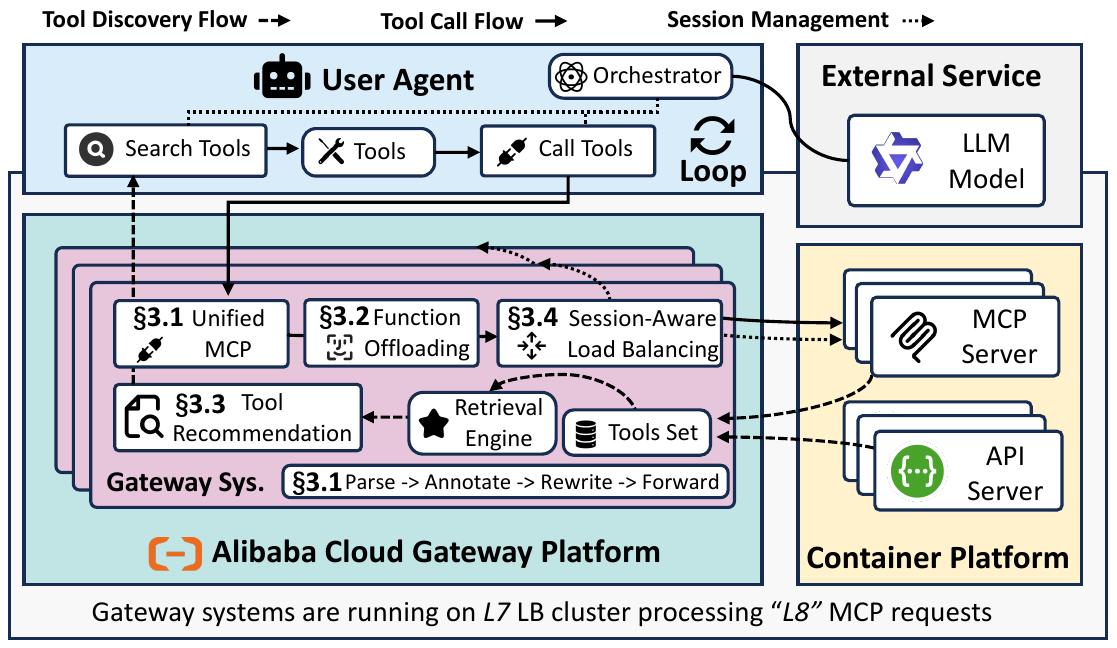}
    \caption{The MCP Gateway architecture}
    \Description{}
    \label{fig:gateway_overview}
\end{figure}

\noindent\textbf{System overview.} The system (Figure~\ref{fig:gateway_overview}) comprises four building blocks across the lifecycle of agent--MCP interaction.
The gateway exposes heterogeneous backends, e.g., MCP transport variants and legacy APIs, through a unified MCP interface, so that a client connects to the gateway as a single MCP endpoint (\S\ref{sec:design_protocol_adaptation}), addressing Challenge~1 and the transport incompatibility in Challenge~2.
On this unified path, the gateway authenticates the client and applies identity-based access control at ingress, and authenticates to backends on behalf of the client at egress, offloading these common functions from backends (\S\ref{sec:design_auth_offload}), addressing the authentication incompatibility in Challenge~2.
Instead of receiving thousands of tools, the agent discovers tools iteratively via the tool recommendation interface, which returns a small set of intent-relevant tools (\S\ref{sec:design_tool_recommendation}), addressing Challenge~3.
Subsequent tool-calling requests are routed to backends via session-aware load balancing. 
Session state is synchronized across gateway replicas, so any gateway instance makes consistent forwarding decisions, preserving session context under replication (\S\ref{sec:design_mcp_lb}), addressing Challenge~4.


\subsection{Protocol Adaptation for Heterogeneity}\label{sec:design_protocol_adaptation}

Tool backends are heterogeneous at both MCP layers (Challenges~1 and~2): at the data layer, legacy OpenAPI services do not implement MCP primitives; at the transport layer, backends diverge across conventional HTTP, HTTP+SSE, and Streamable HTTP.
The gateway therefore provides a unified protocol abstraction across both layers.
\textbf{(1) API protocol conversion.} To onboard legacy APIs as MCP tools at scale, the gateway performs adaptation at both layers.
At the data layer, it translates MCP primitives (e.g., \textit{List Tools} and \textit{Call Tool}) into API invocations and normalizes responses to the MCP tool interface.
Each OpenAPI operation is compiled into exactly one MCP tool.
This one-to-one mapping minimizes tool granularity and reduces ambiguity in tool selection and argument generation.
At the transport layer, it terminates the client-facing MCP connection and communicates with each backend using its required protocol.
\textbf{(2) MCP transport bridging.}
As MCP transport variants and versions are not strictly compatible, the gateway bridges between transport variants (e.g., HTTP+SSE and Streamable HTTP), enabling interoperability without changes to applications.
\textbf{(3) Decoupled adaptation layers.}
The gateway decouples transport bridging from data-layer mapping, rather than coupling them in a monolithic translator.
Since MCP transports evolve rapidly (\S\ref{sec:challenges_of_deploying_mcp}), this decoupling confines protocol churn to the transport stage: new transport variants can be supported without modifying data-layer conversion.

\subsection{Function Offloading} \label{sec:design_auth_offload}
In production deployments, existing tool backends often duplicate auxiliary logic (e.g., authentication and identity-based access control), which increases engineering cost and results in inconsistent enforcement across heterogeneous services.
We therefore adopt gateway-side function offloading: the MCP Gateway provides these shared functions as common infrastructure and enforces them on the end-to-end request path.
Via ingress and egress authentication (\Cref{fig:gateway_overview}), the gateway offloads the following functions:
\textbf{(1) Centralized security enforcement.}
The gateway exposes a unified authentication interface on ingress and maps authenticated identities to backend-specific credentials on egress.
For backends without native authentication, it enforces access policies at ingress and forwards requests over a protected gateway--backend path, establishing an end-to-end security boundary without backend changes.
\textbf{(2) Fine-grained tool access control.}
The gateway determines tool visibility and invocation permissions as a function of identity, enabling policy-driven tool exposure in multi-user MCP services.
Overall, this design reduces backend complexity and ensures consistent enforcement across diverse tool infrastructures.

\subsection{Intent-Driven Tool Recommendation}\label{sec:design_tool_recommendation}

To avoid exposing the full tool catalog to the LLM context (Challenge~3), the gateway supports \emph{intent-driven tool recommendation} (\Cref{fig:gateway_overview}): given the intent (e.g., task description or query context), it returns a small set of relevant tools from the massive backends.
\textbf{(1)~Dual-path indexing.}
Cloud tools are described by both natural language and specialized identifiers (e.g., product and resource names); neither semantic retrieval nor lexical matching alone covers both.
The gateway therefore adopts a semantic--lexical dual-path index, supporting fuzzy intent understanding while preserving precise keyword matching.
\textbf{(2)~Iterative discovery.}
Tool discovery is exposed as an agent-invokable interface. 
The agent can search iteratively in a loop, with or w/o LLM reasoning.
\textbf{(3)~Deterministic and efficient execution.}
Instead of relying on an LLM, tool recommendation runs gateway-side as deterministic, lightweight processing, providing bounded latency under high concurrency and large tool sets.
It is a clear intelligence boundary: reasoning stays on the agent side, while the gateway performs deterministic retrieval.

\subsection{Session-Aware Load Balancing}\label{sec:design_mcp_lb}
Routing requests from the same session to different backends can violate correctness (Challenge~4).
At production scale, the gateway must also sustain high availability under horizontal scaling and failover, which makes session-aware routing a first-order requirement.
Accordingly, the gateway provides session-aware load balancing with three objectives.
\textbf{(1) Session affinity.}
Requests within the same logical session are consistently routed to the same backend MCP server instance under replication.
\revise{
\textbf{(2) Session consistency.}
Under a replicated gateway deployment, different gateway instances must apply the same session-to-backend mapping, which consistent hashing~\cite{karger1997consistent} alone cannot provide. 
This requires consistent routing state to be shared across gateway instances.}
\textbf{(3) Transparency.}
Session-aware routing is enforced within the gateway and is transparent to both MCP clients and backend servers.
These designs preserve session correctness while retaining the scalability and availability benefits of horizontal scaling and failover.
%



\vspace{-0.5em}

\section{Implementation}
We implement the MCP Gateway on Alibaba Cloud's production Layer-7 load balancing infrastructure: protocol processing and session-aware routing are implemented natively in Go ($\sim$8K LoC), the OpenAPI conversion component is compiled into a WebAssembly module ($\sim$6K LoC) loaded by the gateway runtime, and authentication reuses existing L7 load balancer components.
The following subsections describe how the four design blocks in \S\ref{sec:system_design} are realized.

\vspace{-0.5em}

\subsection{Cross-Layer Protocol Translation}
\label{sec:protocol_conversion_routing}

Unlike conventional L7 traffic, which is routed on hosts, paths, and headers, MCP routing is body-driven. 
Tool identity and arguments are carried in JSON-RPC messages within the HTTP body, while session identifiers are transport dependent, appearing in headers or SSE streams.
The gateway therefore performs deep body or stream inspection to enable tool-level routing and track session context.

\noindent\textbf{Protocol conversion for API and MCP.}
Following the one-to-one mapping policy (\S\ref{sec:design_protocol_adaptation}), the gateway implements an automatic OpenAPI-to-MCP compilation pipeline.
At tool mount time, the gateway parses the OpenAPI specification and compiles each operation into an MCP tool definition, generating MCP-compliant metadata (names, descriptions, and schemas) from OpenAPI parameters.
As a result, legacy API operations become invocable through standard MCP primitives (e.g., \textit{List Tools} and \textit{Call Tool}).
The pipeline applies a deterministic rule-mapping, compiled and constraint-checked at mount time (fail-fast); specs lacking semantics are enriched by an offline LLM pass, keeping the serving path deterministic.

\noindent\textbf{Multi-stage request routing.}
MCP services use HTTP+SSE or Streamable HTTP, while legacy OpenAPI services are exposed as MCP tools via protocol conversion.
Upon receiving a request, the gateway resolves the execution target and transport through a staged routing pipeline that \textbf{parses} the tool invocation request in both header and body, \textbf{annotates} it with backend/transport metadata, \textbf{rewrites} it into a backend operation (e.g., protocol conversion for API and MCP), and \textbf{forwards} it to the selected backend.
The gateway terminates client connections and (re-)establishes backend connections according to the resolved transport, keeping the MCP client thin as transports evolve.
This staged structure realizes the decoupling principle in \S\ref{sec:design_protocol_adaptation}: new transports are supported by extending individual routing stages, without changing data-layer conversion or end-to-end integration.
%

\subsection{Authentication and Tool Access Control}
\label{sec:auth_fine_grained}

\begin{figure}[t]
  \centering
  \includegraphics[width=0.76\linewidth]{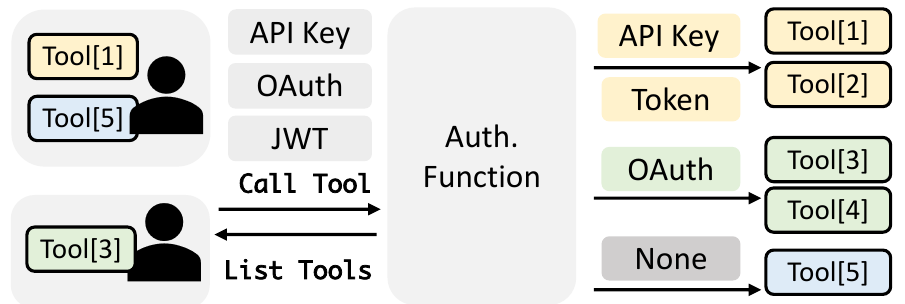}
  \caption{
    Fine-grained access-control function offloading
  }
  \Description{}
  \label{fig:fine_grained_auth}
\end{figure}

As an instantiation of function offloading (\S\ref{sec:design_auth_offload}), the gateway currently offloads two security functions from backends: authentication and fine-grained tool access control (\Cref{fig:fine_grained_auth}).


\noindent\textbf{Authentication with ingress/egress management.}
Each client and backend uses heterogeneous credential schemes; pairing them directly would reintroduce an $N \times M$ integration problem at the credential level.
The gateway implements a two-stage model.
On ingress, it authenticates client requests and establishes a stable identity using standard methods (e.g., API keys, OAuth, and JWT).
On egress, it integrates backend-specific authentication requirements by mapping the established identity to the credentials and formats expected by each backend (e.g., API keys, tokens, or OAuth-based flows).
The gateway provides a security boundary by restricting backend access to gateway-managed channels at ingress.

\noindent\textbf{Fine-grained tool access control.}
The gateway enforces identity-based, tool-level authorization consistently across heterogeneous backends by centralizing authentication management.
Specifically, it grants the access of \textit{List Tools}, \textit{Call Tool}, and the meta tool \textit{Tool Search} (discussed in \S\ref{sec:hybrid_tool_recommendation}) to each authenticated user, ensuring that each authenticated user can only discover and invoke tools permitted by policy.

\subsection{Hybrid Tool Recommendation}
\label{sec:hybrid_tool_recommendation}

The gateway realizes intent-driven recommendation (\S\ref{sec:design_tool_recommendation}) by selecting intent-relevant candidates with deterministic execution; the host then invokes tools through the standard LLM loop over this reduced set.

\noindent\textbf{Interface for tool recommendation.}
To realize agent-driven discovery, the recommendation capability must be invocable from the agent tool-calling loop.
The original MCP \textit{List Tools} does not support query- or constraint-based input as retrieval parameters, limiting the functionality of tool recommendation interfaces.
Extending \textit{List Tools} with additional parameters would reduce compatibility and lead to inconsistent behavior across MCP implementations.
To maximize ecosystem compatibility while enabling tool recommendation, the gateway exposes a \textit{Tool Search} interface as a standard MCP tool.
This tool is always included in the tool list returned by \textit{List Tools} primitive, allowing the host to \textit{discover tools by calling a tool}.
The host submits a \texttt{query} via \textit{Call Tool} and receives a subset of relevant tools in response, with the same structure as the tools returned by \textit{List Tools}.


\begin{figure}[t]
    \centering
    \includegraphics[width=0.8\linewidth]{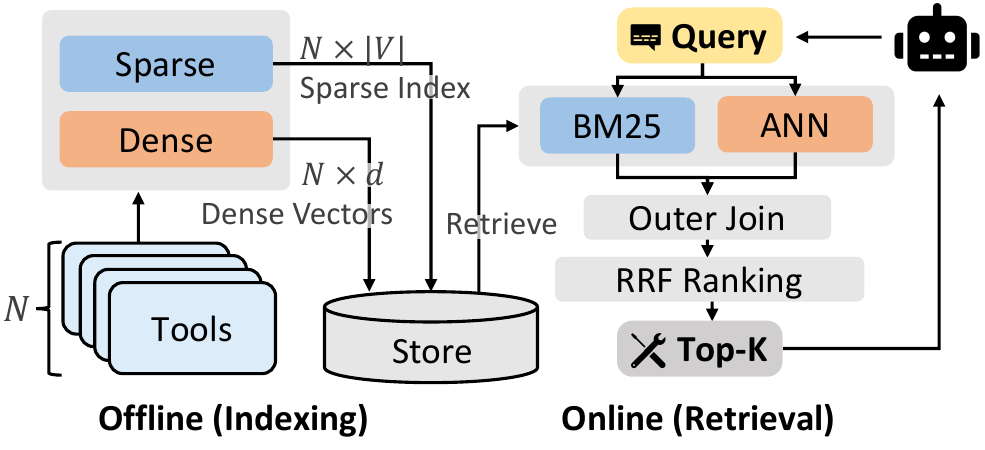}
    \caption{
        Hybrid tool retrieval engine
    }
    \label{fig:tool_search}
\end{figure}


\noindent\textbf{Semantic and lexical indexing and matching.}
The tool retrieval engine incorporates offline indexing and online matching (\Cref{fig:tool_search}).
Offline, at tool mount time, the gateway extracts textual fields (e.g., tool name and description) and constructs two complementary indexes:
(1) a dense index that embeds tool texts into $d$-dimensional vectors with a pretrained encoder, enhancing recall for paraphrases and fuzzy intents;
(2) a sparse full-text index over the tool vocabulary, strengthening exact matching for tool names, resource identifiers, and parameter terms.
At query time, the two paths retrieve candidates independently, via approximate nearest-neighbor (ANN) search on the dense index and BM25 scoring on the sparse index.
The engine then outer-joins the two candidate lists and fuses them with Reciprocal Rank Fusion (RRF), scoring each tool by the sum of its reciprocal ranks with a smoothing constant, and returns the top-$K$ tools.

\begin{reviseblock}
    
Formally, each retrieval path returns $L$ candidates ($L{=}120$). 
A tool $t$ in the union is scored by
\begin{equation}
\mathrm{score}(t)=\sum_{i\in\{\mathrm{dense},\,\mathrm{sparse}\}}\frac{\mathds{1}[t\in C_i]}{\mathrm{RRF}_k+\mathrm{rank}_i(t)}
\label{eq:rrf}
\end{equation}
\noindent where $\mathrm{rank}_i(t)$ is the rank of $t$ in $C_i$, $\mathrm{RRF}_k{=}20$ is the smoothing constant.
The dense path uses cosine similarity; the sparse path uses BM25 ($k_1{=}1.2$, $b{=}0.75$). 
Both the dense and sparse indexes are constructed from tool names, descriptions, and top-level parameter names; the sparse path also splits camelCase identifiers and removes stop words.

\end{reviseblock}

\subsection{Stateful Request Routing}
\label{sec:lb_segmented_consistency}

\begin{figure}[!t]
    \centering
    \includegraphics[width=0.90\linewidth]{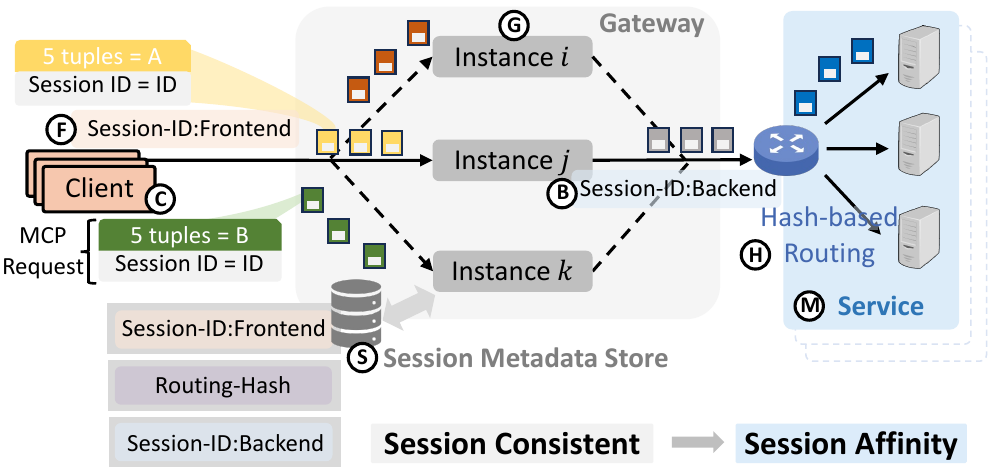}
    \caption{Session-aware request routing}
    \label{fig:segmented_session_consistent}
\end{figure}

\noindent\textbf{Elastic scaling and session-aware routing.}
MCP servers are deployed as multiple replicas behind a service endpoint, and the gateway itself runs as multiple instances.
For stateless deployments, each request can be routed independently to any healthy replica.
The rest of this subsection describes how the gateway enforces session affinity and consistency (\S\ref{sec:design_mcp_lb}) for stateful deployments (\Cref{fig:segmented_session_consistent}).

\noindent\textbf{State synchronization via centralized store.}
A single logical session spans two identifiers: \texttt{Session-ID:\allowbreak Frontend}, generated by the gateway toward the client, and \texttt{Session-ID:\allowbreak Backend} (\circled{B}), assigned by the backend MCP Service.
Any gateway instance serving the session must resolve this mapping consistently.
The gateway stores the mapping in a centralized \emph{Session Metadata Store} (\circled{S} in \Cref{fig:segmented_session_consistent}).

The MCP Gateway (\circled{G}) terminates client (\circled{C}) connections and establishes backend connections to MCP Services (\circled{M}) on demand.
When a client session is created, the gateway generates a \texttt{Session-ID:\allowbreak Frontend}, caches it locally, and creates a record in the Session Metadata Store (\circled{S}).
When subsequent requests trigger access to a specific MCP Service, the gateway establishes the backend session on demand, obtains the \texttt{Session-ID: Backend}, and writes/updates the mapping between \texttt{Session-ID: Frontend} and \texttt{Session-ID: Backend} in the Session Metadata Store.
If later requests of the same session are load-balanced to a different gateway instance, the instance can retrieve the mapping from the store and continue using the backend session, preserving session consistency during scaling events or failover, without any point-to-point synchronization among instances.


\noindent\textbf{Hash-based session affinity in segmented deployments.}
The gateway and MCP services are operated and scaled independently as two tiers; 
the gateway does not track instance churn inside each service.
Since the frontend session is established before the backend session, the gateway computes a \texttt{Routing-Hash} after generating \texttt{Session-ID:Frontend}, and places it into an HTTP request header on the \textit{Gateway} -> \textit{Service} path as the key for intra-service forwarding.
The gateway system uses a hash instead of propagating the original session identifier to hide the raw identifier across the gateway--compute boundary and reduce coupling.
\revise{\texttt{Routing-Hash} is stored as part of the session metadata, together with \texttt{Session-ID:Frontend} and its corresponding \texttt{Session-ID:Backend}.
This allows any gateway instance that later takes over the session to reuse the same routing key, preserving session affinity and cross-gateway consistency across requests.
Each MCP Service exposes a single domain endpoint and applies stateless consistent hashing~\cite{karger1997consistent} to the \textit{Routing-Hash} (\circled{H}) inside the service, mapping requests stably to a specific backend MCP server.
Only a small fraction of sessions should be re-established by the gateway when the backend service scales or fails.}

\begin{figure*}[!tbp]
  \centering
  \begin{subfigure}[b]{0.245\textwidth}
    \centering
    \includegraphics[width=\linewidth]{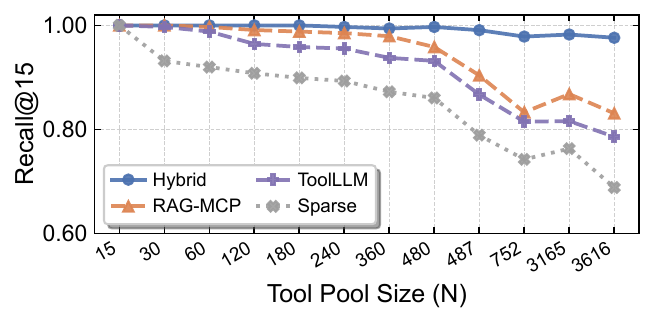}
    \caption{Production: Recall@15}
    \label{fig:cmp-ali-ret}
  \end{subfigure}
  \hfill
  \begin{subfigure}[b]{0.245\textwidth}
    \centering
    \includegraphics[width=\linewidth]{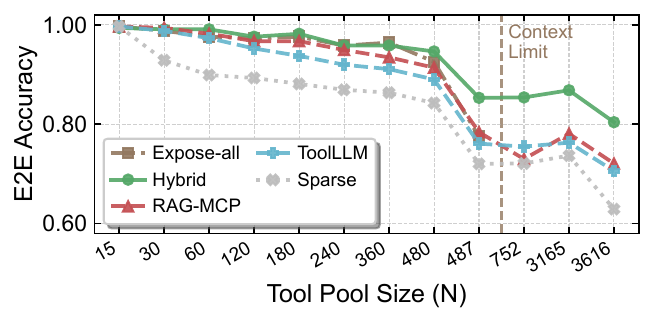}
    \caption{Production: Selection Accuracy}
    \label{fig:cmp-ali-e2e}
  \end{subfigure}
  \hfill
  \begin{subfigure}[b]{0.245\textwidth}
    \centering
    \includegraphics[width=\linewidth]{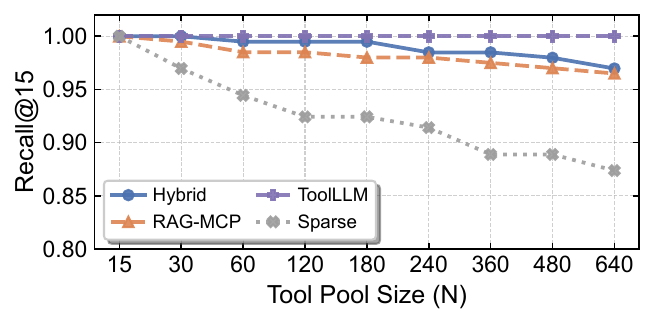}
    \caption{ToolBench: Recall@15}
    \label{fig:cmp-tb-ret}
  \end{subfigure}
  \hfill
  \begin{subfigure}[b]{0.245\textwidth}
    \centering
    \includegraphics[width=\linewidth]{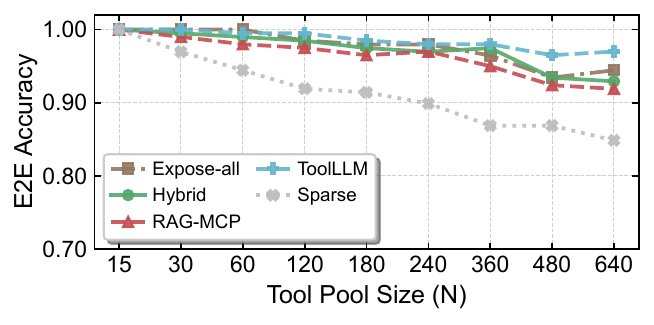}
    \caption{ToolBench: Selection Accuracy}
    \label{fig:cmp-tb-e2e}
  \end{subfigure}
  \caption{Retrieval comparison across methods on production cloud tools (a--b) and ToolBench (c--d), Qwen3.6-plus }
  \label{fig:retrieval-comparison}
\end{figure*}

\noindent\textbf{Long-lived connection bridging with Pub/Sub.}
Different transport modes impose different constraints on session consistency.
For Streamable HTTP, connections are request-scoped, even with streaming, the channel terminates after the response completes, so a reply can always be returned by the gateway instance that received the request.
In contrast, under HTTP+SSE, the client establishes a long-lived, instance-bound SSE channel to a specific gateway instance; all server-to-client events must be delivered through that channel.
This creates a mismatch in a multi-instance gateway, as subsequent requests carrying the same \texttt{Session-ID:Frontend} may be load-balanced to a different gateway instance that does not own the SSE return path, and thus cannot deliver responses to the client.

To resolve this, we implement a lightweight cross-instance coordination layer based on a centralized {\emph{Pub/Sub}} channel.
The key idea is to explicitly separate \emph{where a request is processed} from \emph{where a response must be returned}.
Upon receiving a request, the gateway determines whether the local instance owns the required return path; if not, it forwards either the \emph{request} or the \emph{response} across instances via Pub/Sub.
Specifically, when the frontend uses SSE, requests are redirected to the gateway instance that holds the client--gateway SSE channel; when the backend uses SSE, responses are redirected to the instance that must return the HTTP response.
This bridging is activated only when a cross-instance path discontinuity would otherwise occur, in all other cases, requests and responses are handled locally.
%



\section{Deployment}
\label{sec:deployment}

\noindent\textbf{Production Scale.} MCP Gateway has operated continuously in production for more than 11 months across more than five Alibaba Cloud regions.
\revise{Over this period, tens of thousands of gateways have been deployed.
Each gateway runs as multiple instances and scales out as its workload grows, so the total number of instances is several times the number of gateways.
Session state is synchronized only among the instances of the same gateway; the overhead thus depends on a gateway's instance count, not on the total deployment size.}
\revise{The OpenAPI conversion module supports OpenAPI 3.0 and 3.1 and has been validated on more than 20,000 OpenAPI operations across 315 Alibaba Cloud products.}
The largest observed per-user mount reaches thousands of tools.



\section{Evaluation}


\subsection{Evaluation Setup}

\looseness=-1
\noindent\textbf{Environment.}
\revise{
Gateway pods (2 cores, 4\,GB each) run on Intel Xeon Platinum 8369B servers;
the tool retrieval engine runs separately on 4-core, 16\,GB, 50\,GB-storage
machines and scales independently.
We evaluate in two settings.
Production evaluation demonstrates scalability in one of the clusters and tool retrieval on production tool sets.
Controlled evaluation runs  separatly for concurrent session traffic stress testing and for baseline comparison by replaying production workloads, since online traffic lacks ground truth and the baselines cannot serve real tenants.
}
\noindent\textbf{Workloads.}
\emph{Alibaba Cloud}: up to 3{,}616 tools from our production catalog and sampled queries from production traces over VPC, ECS and SLB services.
\emph{ToolBench}~\cite{qin2024toolllm} (up to 640 tools): a public benchmark
for comparability.
\looseness=-1
\noindent\textbf{Baselines.}
(1) \emph{RAG-MCP}~\cite{gan2025rag}: dense embeddings only;
(2) \emph{ToolLLM}~\cite{qin2024toolllm}: retriever fine-tuned on ToolBench;
(3) \emph{BM25}: sparse retrieval (naive tokenizer, no camelCase splitting);
\revise{(4) \emph{Expose-all}: no retrieval, all tools inlined.}
\noindent\textbf{Protocol and Metrics.}
The retriever returns Top-15, from which the LLM selects one, with \texttt{tool\_choice=required}.
We report \emph{Recall@15} (ground truth in candidates,
\S\ref{sec:eval_tool_recommendation}), \emph{selection accuracy} (LLM picks
the target tool, \S\ref{sec:eval_agent_tool_selection}), and completion time
and token usage (\S\ref{sec:eval_agent_tool_selection}). 



\subsection{Agent Tool Selection Performance}
\label{sec:eval_agent_tool_selection}

\noindent\textbf{Tool selection accuracy on cloud-scale tools.}
\revise{Any retrieval and selection pipeline cuts tokens and latency; we first investigate whether it could preserve or improve the accuracy over the expose-all approach.}
\revise{
As shown in Figure~\ref{fig:cmp-ali-e2e} and \ref{fig:cmp-tb-e2e}, the accuracy of the expose-all approach degrades as the tool pool grows:} 
on ToolBench, the selection accuracy drops from 100\% at $N{=}15$ to 94.4\% at $N{=}640$, which costs much more tokens than retrieval-based methods; 
on Alibaba Cloud tools, it exceeds the LLM context window at $N{\geq}752$ and becomes infeasible for benchmarking.
In contrast, retrieval-based methods maintain feasible accuracy, at $N{=}3{,}616$, Hybrid achieves 80.4\% versus 72.1\% for RAG-MCP and 70.6\% for ToolLLM.
\revise{Thus, the hybrid retrieval we applied achieves scalability and benefits as a retrieval method.}

\aptLtoX{\begin{table}[tbp]
\centering
\small
\setlength{\tabcolsep}{6pt}
\renewcommand{\arraystretch}{0.68}
\caption{Tool selection time cost, Qwen3-Plus}
\label{tab:e2e_latency}
\begin{tabular}{lrrrr}
\toprule
& Avg (ms) & P50 (ms) & P95 (ms) & P99 (ms) \\
\midrule
Top-15 & 6236.96 & 5527.63 & 12399.04 & 15666.76 \\
Top-30 & 8310.92 & 7683.10 & 14855.51 & 16344.22 \\
Top-60 & 14364.14 & 14208.26 & 21085.03 & 25581.55 \\
\revise{Expose-all} & 55662.68 & 55029.86 & 59962.32 & 67936.70 \\
\bottomrule
\end{tabular}
\end{table}

\begin{table}[tbp]
\centering
\small
\setlength{\tabcolsep}{6pt}
\renewcommand{\arraystretch}{0.68}
\caption{Token consumption, Qwen3-Plus}
\label{tab:e2e_tokens}
\begin{tabular}{lrrrr}
\toprule
& Avg (k) & P50 (k) & P95 (k) & P99 (k) \\
\midrule
Top-15 & 21.22 & 20.67 & 28.29 & 43.75 \\
Top-30 & 39.54 & 39.68 & 56.48 & 62.57 \\
Top-60 & 79.75 & 81.56 & 105.70 & 121.66 \\
\revise{Expose-all} & 505.86 & 507.29 & 511.38 & 512.56 \\
\bottomrule
\end{tabular}
\end{table}}{\begin{table}[tbp]
\centering
\small
\setlength{\tabcolsep}{6pt}
\renewcommand{\arraystretch}{0.68}

\begin{minipage}{\linewidth}
\centering
\captionof{table}{Tool selection time cost, Qwen3-Plus}
\label{tab:e2e_latency}
\begin{tabular}{lrrrr}
\toprule
 & Avg (ms) & P50 (ms) & P95 (ms) & P99 (ms) \\
\midrule
Top-15   & 6236.96  & 5527.63  & 12399.04 & 15666.76 \\
Top-30   & 8310.92  & 7683.10  & 14855.51 & 16344.22 \\
Top-60   & 14364.14 & 14208.26 & 21085.03 & 25581.55 \\
\revise{Expose-all} & 55662.68 & 55029.86 & 59962.32 & 67936.70 \\
\bottomrule
\end{tabular}
\end{minipage}

\begin{minipage}{\linewidth}
\centering
\captionof{table}{Token consumption, Qwen3-Plus}
\label{tab:e2e_tokens}
\begin{tabular}{lrrrr}
\toprule
 & Avg (k) & P50 (k) & P95 (k) & P99 (k) \\
\midrule
Top-15   & 21.22  & 20.67  & 28.29  & 43.75  \\
Top-30   & 39.54  & 39.68  & 56.48  & 62.57  \\
Top-60   & 79.75  & 81.56  & 105.70 & 121.66 \\
\revise{Expose-all} & 505.86 & 507.29 & 511.38 & 512.56 \\
\bottomrule
\end{tabular}
\end{minipage}
\vspace{-0.5em}
\end{table}}

\begin{figure}[tbp]
  \centering
  \begin{subfigure}[b]{0.45\linewidth}
    \centering
    \includegraphics[width=\linewidth]{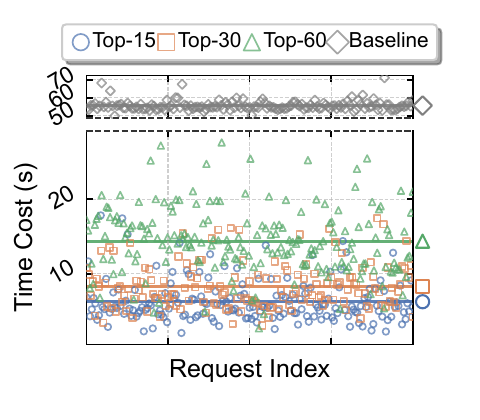}
    \caption{Time cost}
    \label{fig:e2e_latency}
  \end{subfigure}\hfill
  \begin{subfigure}[b]{0.45\linewidth}
    \centering
    \includegraphics[width=\linewidth]{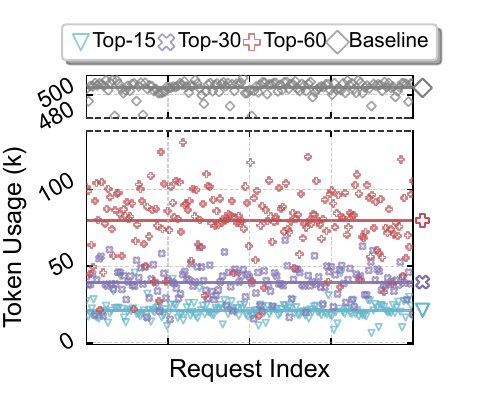}
    \caption{Token usage}
    \label{fig:e2e_token_usage}
  \end{subfigure}
  \caption{LLM agent tool selection performance}
  \label{fig:e2e_performance}
  \Description{}
\end{figure}

\noindent\textbf{Tool selection latency and token cost.}
Figure~\ref{fig:e2e_performance} shows that the end-to-end agent tool use completion evaluation follows the Agent--MCP Gateway--Tools path and uses representative cloud tools and production user traffic (VPC, ECS and SLB tools) as a reference point.
The gateway hosts $O(1k)$ tools for cloud resource operations.
In the expose-all setting, the agent selects the target from 372 tools, since it cannot scale to $O(1k)$ tools.
In the retrieval-based setting, the agent selects tools from the recommended list (Top-15, 30, 60), and then issues tool calls through the MCP Gateway.
As shown in Table~\ref{tab:e2e_latency}, on average, Top-15 reduces time cost from 55.7s (expose-all) to 6.2s, a reduction of $8.9 \times$.
Table~\ref{tab:e2e_tokens} indicates that token consumption decreases from 505.86k to 21.22k, a reduction of $23.8 \times$.
On tail latency, Top-15 reduces P95 from 60.0 s to 12.4 s.\footnotemark{}
\revise{The agent can benefit this only if the retrieval maintains high recall.}

\footnotetext{The reported latency measures the interval from the agent initiating tool recommendation to the gateway receiving the subsequent tool-call request.}



\subsection{Hybrid Tool Recommendation}
\label{sec:eval_tool_recommendation}

\aptLtoX{\begin{figure}[t]
\centering
\includegraphics[width=\linewidth]{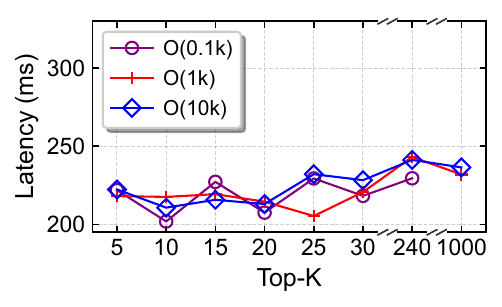}
\caption{Retrieval latency v.s. Tool scale}
\label{fig:latency_vs_tool_scale}
\Description{Retrieval engine latency vs tool scale.}
\end{figure}

\begin{figure}[t]
\centering
\includegraphics[width=\linewidth]{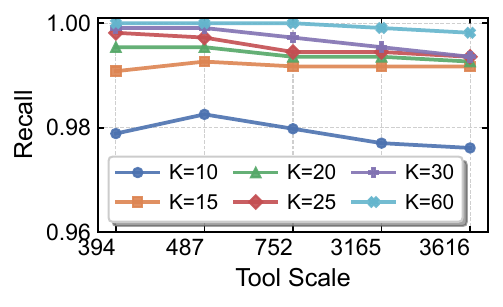}
\caption{Top-K recall v.s. Tool scale}
\label{fig:recall_vs_tool_scale}
\Description{Top-K recall vs tool scale.}
\end{figure}}{\begin{figure}[t]
\centering
\begin{minipage}[t]{0.22\textwidth}
  \centering
  \includegraphics[width=\linewidth]{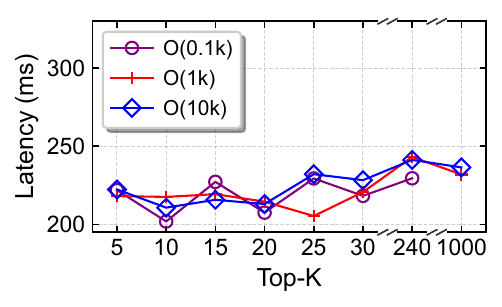}
  \caption{Retrieval latency v.s. Tool scale}
  \label{fig:latency_vs_tool_scale}
\end{minipage}
\begin{minipage}[t]{0.22\textwidth}
  \centering
  \includegraphics[width=\linewidth]{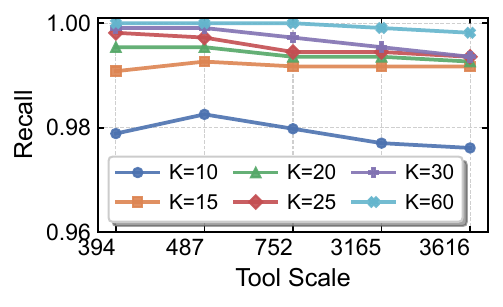}
  \caption{Top-K recall v.s. Tool scale}
  \label{fig:recall_vs_tool_scale}
\end{minipage}
\Description{(Left) Retrieval engine latency vs tool scale. (Right) Top-K recall vs tool scale.}
\end{figure}}

\looseness=-1
\noindent\textbf{Latency and recall at scale.}
Figure~\ref{fig:latency_vs_tool_scale} shows that latency remains stable from $O(0.1k)$ to $O(10k)$ tools.
Across different retrieval result set sizes, i.e., Top-5, Top-10, Top-15, and Top-1000, the retrieval process complete within 202--243 ms, which is bounded.
\revise{
    As illustrated in Figure~\ref{fig:recall_vs_tool_scale}, recall also remains high as the tool set grows.}
Note that, as the mounted tool set increases from 394 to 3616 tools, Top-15 retrieval can still achieve over 98\% recall (larger query sets than Figure~\ref{fig:cmp-ali-ret}).
It further shows a recall gain when expanding the recommendation set from Top-10 to Top-15.
This indicates Top-15 is an effective result set size for tool recommendation. 
Beyond these, our production traces reveal additional types of \emph{weak-semantic} queries (e.g., ID-only requests). 
The semantic signal of these requests is too weak for retrieval. 
We report this observation and production optimization as (Lesson~4) in \S\ref{sec:lessons_learned} .

\looseness=-1
\noindent\textbf{Comparison with retrieval baselines.}
On ToolBench (Figure~\ref{fig:cmp-tb-ret} and \ref{fig:cmp-tb-e2e}), ToolLLM achieves the best recall across all sizes, since ToolLLM's retriever was trained (fine-tuned) over ToolBench datasets.
However, this advantage does not transfer on other domain, if we do not re-tune it. 
On Alibaba Cloud tools (Figure~\ref{fig:cmp-ali-ret} and~\ref{fig:cmp-ali-e2e}), ToolLLM's recall drops as the tool pool grows, falling to 78.6\% at $N{=}3{,}616$, below its performance on ToolBench.
In production, the tool set may change according to tenants continuously, such re-tuning may become impractical, because every update requires data generation, retraining, and redeployment.

Without fine-tuning, the hybrid retrieval consistently outperforms on Alibaba Cloud tools.
At $N{=}3{,}616$, Hybrid maintains $\sim$97.6\% Recall@15 while RAG-MCP (dense-only) drops to 83.1\%, ToolLLM to 78.6\%, and Sparse (BM25) to 68.8\%.
The advantage of combining semantic and lexical matching is good at large scale.
For the cloud tools, dense embeddings capture intent-level information, while BM25 preserves exact matches on resource identifiers and parameter names.
On ToolBench, where tool descriptions are more generic, the hybrid approach still matches or slightly outperforms RAG-MCP (97.0\% vs.\ 96.5\% at $N{=}640$).



\subsection{Gateway Overhead and Scalability}
\label{sec:eval_gateway_overhead_and_scalability}

\begin{figure}[tbp]
        \centering

        \begin{subfigure}[b]{0.32\linewidth}
          \centering
          \includegraphics[width=\linewidth]{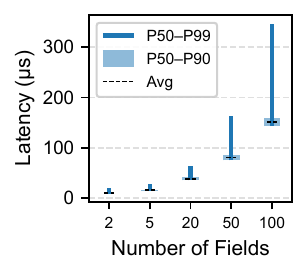}
          \caption{Field count}
          \Description{Conversion latency percentiles by number of fields.}
          \label{fig:conversion_latency_fields}
        \end{subfigure}
        \hfill
        \begin{subfigure}[b]{0.32\linewidth}
          \centering
          \includegraphics[width=\linewidth]{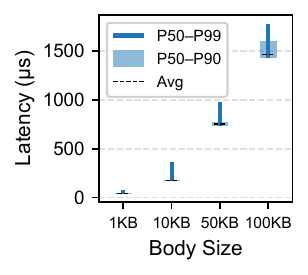}
          \caption{Body size}
          \Description{Conversion latency percentiles by body size.}
          \label{fig:conversion_latency_bodysize}
        \end{subfigure}
        \hfill
        \begin{subfigure}[b]{0.32\linewidth}
          \centering
          \includegraphics[width=\linewidth]{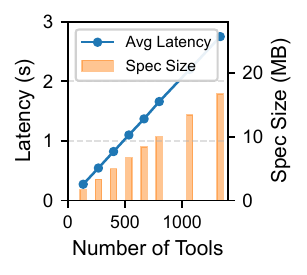}
          \caption{Offline}
          \Description{Offline conversion latency and spec size versus number of tools.}
          \label{fig:conversion_offline}
        \end{subfigure}

        \caption{API conversion overhead: online request conversion (a, b) and offline tool generation (c).}
        \Description{Three plots showing online conversion latency percentiles and offline conversion cost.}
        \label{fig:conversion_latency}
  \end{figure}

\noindent\textbf{API conversion overhead is negligible.}
For the online part, Figure~\ref{fig:conversion_latency} reports the latency of converting an MCP tool call request into the corresponding API invocation on the gateway.
Payloads vary from 2--100 JSON fields and 1--100KB bodies.
As shown in Figure~\ref{fig:conversion_latency_fields}, for typical requests with fewer than 20 fields, the median conversion latency stays below 35\,$\mu$s.
For the largest payload with 100 fields, the P50 latency is 143\,$\mu$s.
Figure~\ref{fig:conversion_latency_bodysize} shows that the conversion scales linearly with body size.
Figure~\ref{fig:conversion_offline} reports the offline conversion that generates MCP tools from API specs.
For a large one of 16.7\,MB, generating 1{,}334 tools takes only 2.75\,s.
The conversion time grows linearly with the number of tools, it runs once per registration, the cost can be amortized.

\begin{figure}[t]
    \centering

    \begin{subfigure}[b]{0.45\linewidth}
      \centering
      \includegraphics[width=\linewidth]{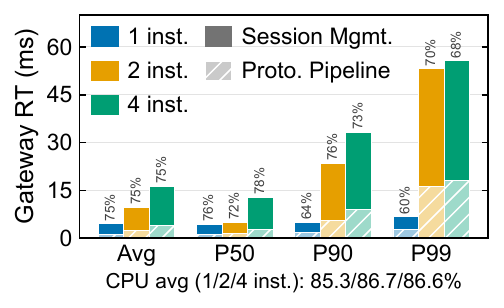}
      \caption{Stateful gateway}
      \Description{}
      \label{fig:gateway_rt_v4_stateful}
    \end{subfigure}
    \hfill
    \begin{subfigure}[b]{0.45\linewidth}
      \centering
      \includegraphics[width=\linewidth]{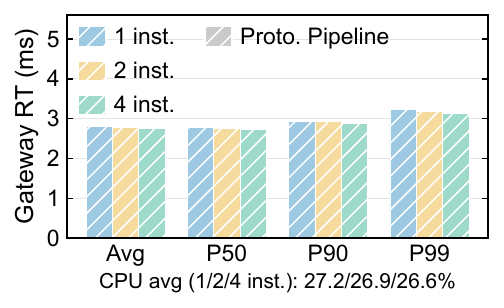}
      \caption{Stateless gateway}
      \Description{}
      \label{fig:gateway_rt_v4_stateless}
    \end{subfigure}

    \caption{Gateway RT breakdown vs.\ \# instances}
    \Description{}
    \label{fig:gateway_rt_v4}
\end{figure}

\noindent\textbf{Session-aware routing dominates per-request cost.}
Figure~\ref{fig:gateway_rt_v4_stateful} shows that stateful routing adds substantial overhead, even before horizontal scale-out.
\revise{With a single gateway instance at $\sim$100 QPS (under stress with concurrent sessions), the mean RT is 4.59 ms with P50/P90/P99 of 4.41/5.02/6.81 ms.}
Session Management---session-affinity lookup and routing---dominates the cost, contributing 3.45 ms on average (75\% of mean RT) and 4.07 ms at P99, while the Protocol Pipeline (transport parsing and MCP dispatch) costs 1.14 ms.
At this load, the stateful instance already consumes 85.3\% CPU, versus 27.2\% for the stateless one.

\noindent\textbf{Cross-instance misses amplify latency.}
\revise{As the gateway scales out, this overhead grows. With four instances, mean RT rises to 16.26 ms and P99 reaches 55.61 ms, about $8\times$ the single-instance P99, with Session Management P99 growing from 4.07 ms to 37.55 ms.}
The increase comes mainly from session data misses. 
When the requests of a session are distributed to an instance that does not hold the session, the gateway must fetch session metadata from the store.
With two instances, P50 latency remains low as the single-instance while P90 increases to 23.58ms, with four instances, the latency rises further and makes P50 to 12.87ms.
We can compare this with stateless ones in (Figure~\ref{fig:gateway_rt_v4_stateless}), across instance counts, mean RT stays within 2.76--2.81 ms, P99 below 3.3 ms, and CPU around 27\%.
While the centralized store enables stateless scale-out, it places the latency bottleneck on the miss path, especially for MCP traffic, where we observe a $2\times$ read/write amplification to the store (\S\ref{sec:lessons_learned}).

\begin{figure}[t]
  \centering
  \begin{subfigure}[b]{0.47\linewidth}
    \centering
    \includegraphics[width=\linewidth]{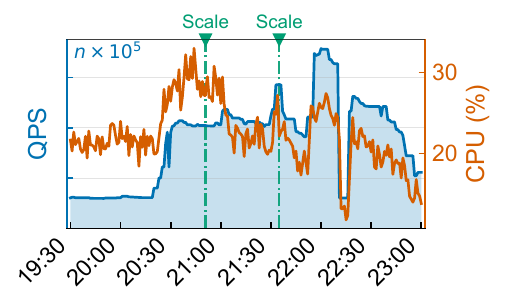}
    \caption{QPS \& CPU usage}
    \label{fig:gw_qps_cpu}
  \end{subfigure}\hfill
  \begin{subfigure}[b]{0.47\linewidth}
    \centering
    \includegraphics[width=\linewidth]{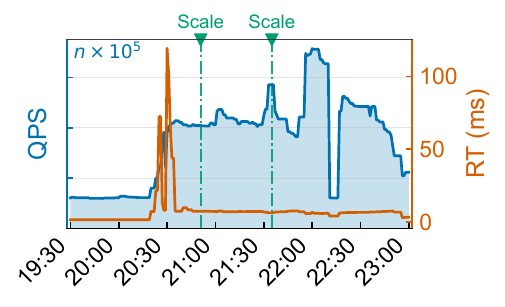}
    \caption{QPS \& RT}
    \label{fig:gw_qps_rt}
  \end{subfigure}
  \caption{The scalability of the MCP Gateway}
  \label{fig:the_scalability_of_the_gateway}
\end{figure}

\noindent\textbf{Elastic scale-out with stable RT.}
\revise{Figure~\ref{fig:the_scalability_of_the_gateway} shows a production trace from one of our largest clusters.
It shows that the gateway maintains stable response time (RT) around 7 ms.}
As load increases, the gateway scales out on CPU pressure, at 20:50:52, after traffic rises around 20:30 and per-instance CPU increases from $\sim$20\% to $\sim$30\% (Figure~\ref{fig:gw_qps_cpu}), the cluster scales and drives CPU back toward the $\sim$20\% steady level without sacrificing RT.
A second scale-out occurs at 21:35:00 under a QPS-driven CPU increase, again with stable RT.
Note that the RT spike around 20:30:00 is due to a backend scale-out failure causing request backlog on the gateway, rather than a limitation of the gateway scaling.

\revise{
    Each instance caches session metadata locally. 
    It accesses the centralized store only when (i) a new session first lands on it, or (ii) a newly added instance first sees an existing session.
    Figure~\ref{fig:gateway_rt_v4} stresses the gateway with concurrent new-session traffic.
    Figure~\ref{fig:the_scalability_of_the_gateway} shows that in production, a session carries more than one request, so the cost of the first miss is amortized over subsequent requests.
    Scale-out adds only a few instances as needed (typically 1--4).
    As a result, RT stays flat across scale-out events.
}




\vspace{-1em}

\section{Lessons Learned from Deployments}\label{sec:lessons_learned}


\begin{figure}[t]
    \centering

    \begin{subfigure}[b]{0.65\linewidth}
      \centering
      \includegraphics[width=\linewidth]{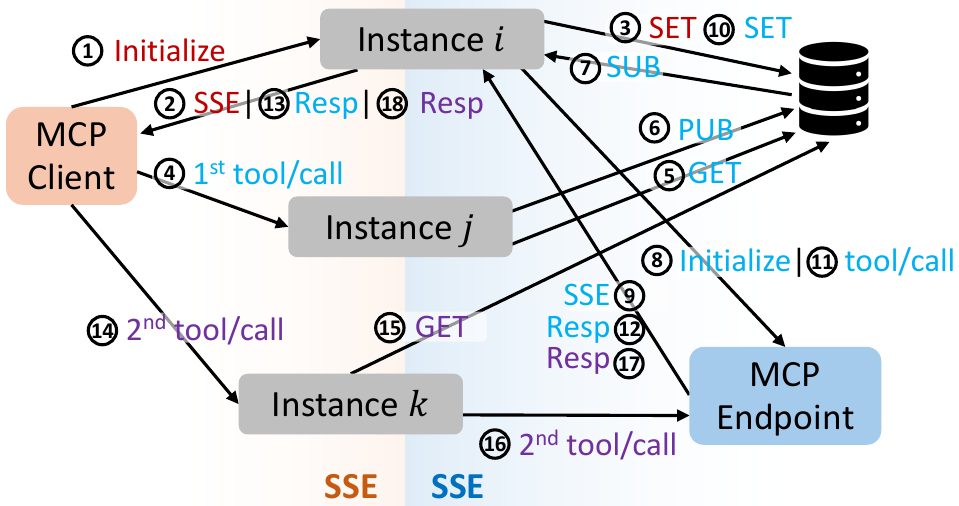}
      \caption{Centralized architecture}
      \label{fig:session_management_badcase}
    \end{subfigure}
    \begin{subfigure}[b]{0.65\linewidth}
      \centering
      \includegraphics[width=\linewidth]{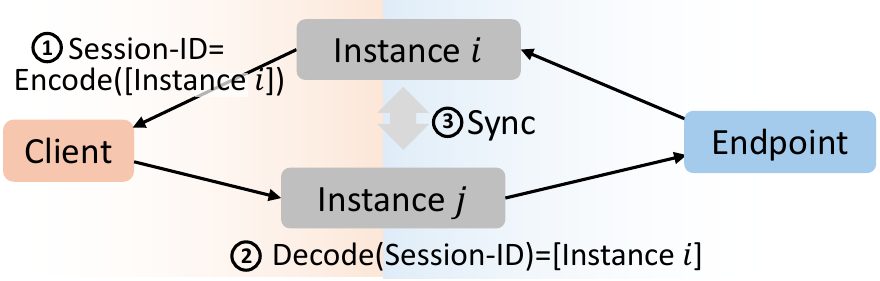}
      \caption{Session-ID encoding architecture}
      \label{fig:session_management_encoding}
    \end{subfigure}


    \caption{From centralized management to encoding.}
    \Description{Illustration of the evolution from centralized session management to Session-ID encoding.}
    \label{fig:session_management_lesson}
\end{figure}



\noindent\textbf{Lesson 1: Coordination for session consistency.}
In production, we observed that when both frontend and backend sessions use SSE, a session is initialized on one gateway instance while subsequent tool calls can be routed to others (Figure~\ref{fig:session_management_badcase}).
To preserve correctness, the deployment relies on a centralized coordination module (e.g., a session store such as Redis plus Pub/Sub).
In the worst case, handling only three MCP requests (one initialization and two \textit{Call Tool} requests) triggers six interactions with the centralized module (\circled{3}\circled{5}\circled{6}\circled{7}\circled{10}\circled{15}), making it both a bottleneck and a reliability risk point.
Moreover, it introduces a hard external dependency, increasing cost and coupling the gateway SLA to the external store and messaging system.
An alternative is fully distributed synchronization among gateway instances, but it is complex and can generate substantial synchronization traffic, especially under elastic scaling.

However, there is a property of MCP traffic, the gateway-assigned \texttt{Session-ID} should be carried by the MCP client on every subsequent request.
This makes the \texttt{Session-ID} an effective signal for routing and coordination. 
Our system is evolving toward a \emph{Session-ID encoding} approach (not yet deployed at scale).
As shown in Figure~\ref{fig:session_management_encoding}, during session establishment, the gateway encodes an owner hint (e.g., an instance identity) into the \texttt{Session-ID} returned to the client.
When a later request of the same session arrives at a different gateway instance, the receiving instance decodes the \texttt{Session-ID} to locate the owner and coordinates with that instance if needed.
Despite being fully distributed, this design keeps synchronization overhead bounded.

\begin{lessonbox}
\textbf{Lesson 1:} \textit{Centralized session stores become the bottleneck at scale; exploit identifiers the protocol already carries---encode ownership into the Session-ID.}
\end{lessonbox}


\begin{figure}[!t]
  \centering
    \includegraphics[width=\linewidth]{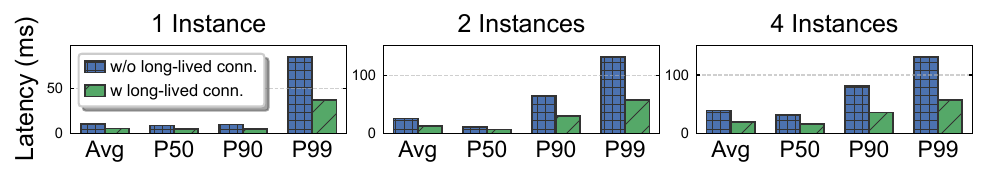}
    \caption{Latency w/o and with long-lived connection}
  \label{fig:latency_comparison_long_lived_connection}
\end{figure}


\noindent\textbf{Lesson 2: Long-lived gateway--backend connection.}
A naive design was initially adopted to simplify implementation:
the gateway did not maintain persistent connections to the backend. 
Each tool call triggered a full connect--initialize--invoke sequence, followed by session and connection termination.
Deployment measurements show that this design introduces substantial user-aware latency. 
From the client perspective, a tool call is a single request, but the naive design expands it into three gateway--backend interactions. 
As shown in Figure~\ref{fig:latency_comparison_long_lived_connection}, without long-lived gateway--backend connections, the 4-instance tail latency exceeds 100ms.
The  lesson is that, when the backend requires long-lived connections and sessions, the gateway should absorb this complexity to avoid overhead on the critical path. 
After the redesign, long-lived gateway--backend connections reduce mean/P90/P99 latency by up to 51.7\%/55.7\%/56.5\%.

\begin{lessonbox}
\textbf{Lesson 2:} \textit{When backends require long-lived connections, the gateway should absorb that complexity.}
\end{lessonbox}


\begin{figure}[!t]
  \centering
  \includegraphics[width=0.8\linewidth]{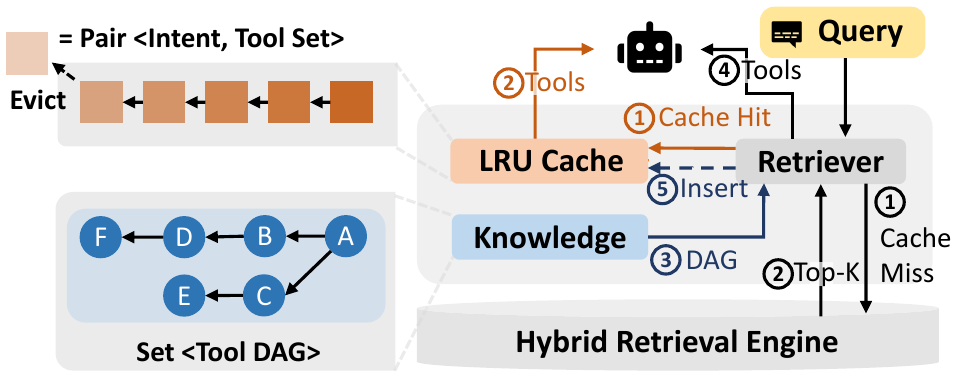}
  \caption{
    Semantic cache
  }
  \Description{}
  \label{fig:semantic_cache}
\end{figure}



\begin{figure}[tp]
    \centering
    \begin{subfigure}[b]{0.22\textwidth}
      \centering
      \includegraphics[width=\linewidth]{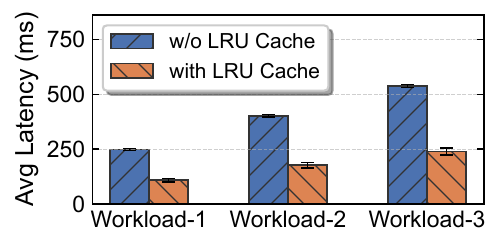}
      \caption{}
      \label{fig:lru_cache_search_time}
    \end{subfigure}
    \begin{subfigure}[b]{0.22\textwidth}
      \centering
      \includegraphics[width=\linewidth]{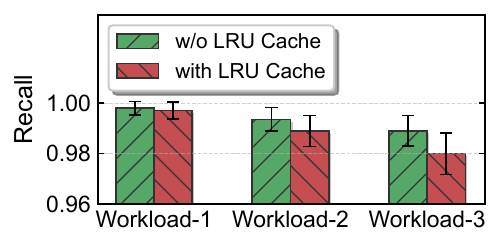}
      \caption{}
      \label{fig:lru_cache_recall}
    \end{subfigure}

    \caption{LRU caching on latency with varied production trace workloads (a); and recall with different query sets (b)}
    \Description{}
    \label{fig:plot-semantic-cache}
  \end{figure}



\begin{table}[t]
  \vspace{0.5em}
  \centering
  \renewcommand{\arraystretch}{0.75}
  
  \caption{Knowledge Cache evaluation.
    Prereq.\ Completeness measures the fraction of queries
    whose \emph{all} prerequisite tools are retrieved.
    Extra RTTs and Tokens are per-query averages incurred by
    iterative LLM discovery.}
  \label{tab:kcache-summary}
  \small
  \begin{tabular}{@{}ll rrr@{}}
    \toprule
    \textbf{Set} & \textbf{Method}
      & \textbf{Compl.(\%)} & \textbf{+RTTs} & \textbf{+Tokens} \\
    \midrule
    \multirow{3}{*}{\shortstack[l]{Set-1\\(394 tools)}}
      & Baseline        & 29.1 & 0    & 0   \\
      & Discovery       & 73.3 & 1.03 & 114 \\
      & Know.\ Cache    & \textbf{100.0} & \textbf{0} & \textbf{0} \\
    \midrule
    \multirow{3}{*}{\shortstack[l]{Set-2\\(3616 tools)}}
      & Baseline        & 24.5 & 0    & 0   \\
      & Discovery       & 56.1 & 1.00 & 110 \\
      & Know.\ Cache    & \textbf{100.0} & \textbf{0} & \textbf{0} \\
    \bottomrule
  \end{tabular}
  \vspace{0.5em}
\end{table}


\noindent\textbf{Lesson 3: Semantic caching for tool recommendation.}
Users often describe an \emph{end goal} in natural language, while successful execution often requires a tool-calling chain with strict prerequisites.
For instance, creating an Application Load Balancer typically requires creating and configuring a VPC and VSwitch first.
This exposes a limitation of semantic-similarity-only tool recommendation, it may retrieve the target tool but miss its prerequisite tools.
When prerequisites are missing, the agent-side LLM must discover them at inference time by issuing additional retrieval-tool calls, each incurring an extra LLM round-trip, added latency, and token overhead.
As dependency chains grow deeper, this iterative discovery compounds, each step has an independent success rate below or around 90\%, so the probability of recovering the full chain decays exponentially with depth.

\noindent (1) Knowledge Cache.
To eliminate the above failure, the MCP Gateway introduces a \emph{Knowledge Cache} that precomputes prerequisite chains from the tool schemas (Figure~\ref{fig:semantic_cache}).
At indexing time, the gateway analyzes dependencies across the tool corpus and builds tool chain DAGs via topological sort.
At query time, after the retrieval engine returns candidate tools, the Retriever looks up each candidate's prerequisite chain in the Knowledge Cache and combines the prerequisite tools into the final result, so the agent receives an executable chain in one shot, with no additional LLM round-trips needed.
\revise{
As shown in Table~\ref{tab:kcache-summary}, the Knowledge Cache achieves 100\% prerequisite completeness on both sets with zero extra RTTs and zero extra tokens, while iterative LLM discovery requires ${\sim}$1 extra round-trip per query on average. 
Also, the success rate can drop to 0\% completeness at chain depth~$\geq$5 in our production measurements.
}

\looseness=-1
\noindent (2) LRU Cache. We observed that online queries exhibit strong semantic skewness, where a small set of similar queries dominates.
We cluster the query intents in a representative production trace into 224 clusters, and the top 10 clusters account for 80.2\% of the traffic.
The MCP Gateway adds an LRU \emph{semantic cache} for $\langle \textit{Intent}, \textit{Tool Set} \rangle$ pairs to bypass external retrieval on cache hits.
Even a small cache (e.g., 64 entries) speeds up retrieval by up to 2.29$\times$ under varied workloads with negligible recall loss (only related to the semantic characteristics of query workloads) (Figure~\ref{fig:lru_cache_search_time}, Figure~\ref{fig:lru_cache_recall}).

\begin{lessonbox}
\textbf{Lesson 3:} \textit{Semantic alone cannot recover tool dependencies; precompute prerequisite chains offline and cache skewed intents for low latency.}
\end{lessonbox}

\looseness=-1
\noindent\textbf{Lesson 4: Query rewriting for domain-specific, under-specified intents.}
In practice, many user queries are weak at semantic.
A common case is an \emph{ID-only} query that contains just an identifier string (e.g., an instance or user ID).
Such inputs carry little semantic signal, while tool descriptions are in natural language.
This mismatch destabilizes retrieval-based tool recommendation, as the system cannot reliably infer the target entity or intended action from an identifier alone.
To address the mismatch, as shown in Figure~\ref{fig:recall_by_optimizations}, a sequence of optimizations is introduced, improving Top-15 Recall from 80.8\% (O0) to 82.4\% (O1), 99.5\% (O2), and further to 99.7\%.
%


\begin{figure}[!t]
    \centering
    \includegraphics[width=0.75\linewidth]{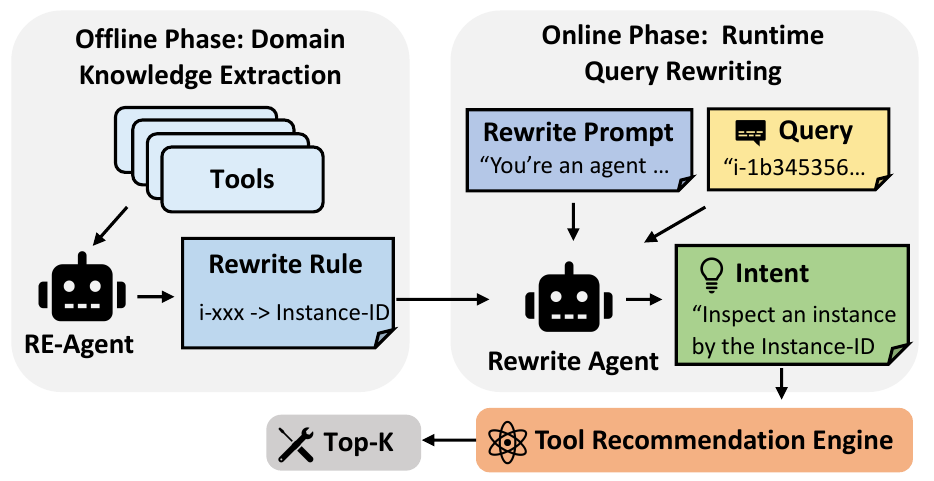}
    \caption{
        Rewrite-Agent and RE-Agent workflow
    }
    \label{fig:re_agent}
\end{figure}



\aptLtoX{\begin{figure}[!t]
\centering
\includegraphics[width=0.235\textwidth]{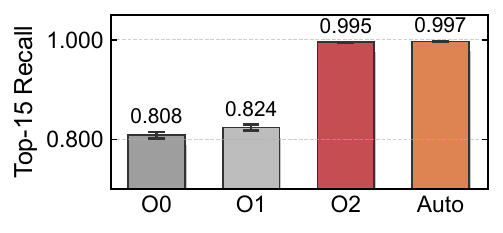}
\caption{Query rewrite optimizations (on Trace-3)}
\Description{}
\label{fig:recall_by_optimizations}
\end{figure}

\begin{figure}[!t]
\centering
\includegraphics[width=0.235\textwidth]{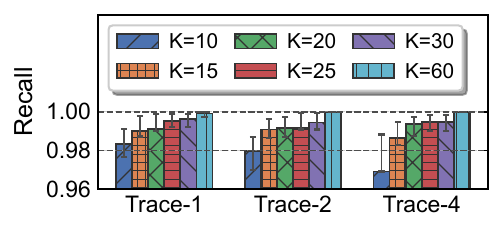}
\caption{Top-K recall v.s. Traces}
\Description{}
\label{fig:topk_recall_vs_traces}
\end{figure}}{\begin{figure}[!t]
\centering
\begin{minipage}[t]{0.235\textwidth}
  \centering
  \includegraphics[width=\linewidth]{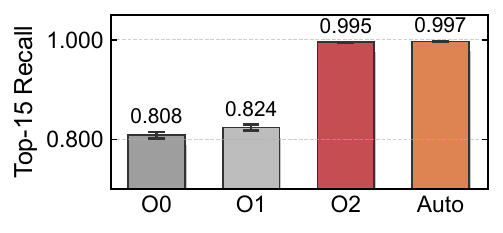}
  \caption{Query rewrite optimizations (on Trace-3)}
  \Description{}
  \label{fig:recall_by_optimizations}
\end{minipage}\hfill
\begin{minipage}[t]{0.235\textwidth}
  \centering
  \includegraphics[width=\linewidth]{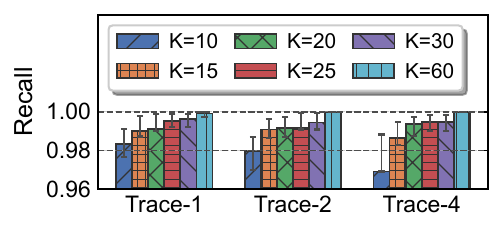}
  \caption{Top-K recall v.s. Traces}
  \Description{}
  \label{fig:topk_recall_vs_traces}
\end{minipage}
\end{figure}}


%
We first implement an LLM-based \emph{Rewrite-Agent} to rewrite the input into a normalized, domain-aware intent before tool recommendation (Figure~\ref{fig:re_agent}). 
Given a fixed prompt template, it converts the raw query into an \emph{Intent} that better matches tool descriptions, and the retrieval engine then recommends the Top-K tools based on the intent (O1 in Figure~\ref{fig:recall_by_optimizations}). 
The key is \emph{entity normalization}, mapping identifiers and other non-semantic strings to an abstract entity type used in tool descriptions (e.g., rewriting an instance ID into an intent targeting a \textit{compute instance}).

However, a general Rewrite-Agent often performs poorly for ID-only queries because it lacks domain knowledge.
Identifier formats, object taxonomies, and naming conventions vary across services, and the model may map an identifier to the wrong entity type, which directly degrades recall.
To address this, we first encode domain knowledge as a set of rewrite rules in the Rewrite prompt.
These rules map common identifier patterns to entity types, steering the Rewrite-Agent to produce intents that align with tool descriptions.
With rule-guided rewriting, the Top-15 recall of tool recommendation increases to 99.5\% (O2 in Figure~\ref{fig:recall_by_optimizations}).

Maintaining such rules manually does not scale.
We therefore design an \emph{RE-Agent} (Rule-extraction agent, Figure~\ref{fig:re_agent}) that automatically extracts rewrite rules from tool descriptions.
Each rule maps an identifier pattern to an entity type.
We inject the extracted rules into the Rewrite prompt to guide normalization, and then let the Rewrite-Agent produce the final Intent.
This automated, rule-guided pipeline improves tool recommendation for ID-only and weak-semantic queries and maintains Top-15 recall of 99.7\% (Auto in Figure~\ref{fig:recall_by_optimizations}).

Using the final configuration (Auto), we analyze multiple production traces where the gateway serves $O(1k)$ cloud operation tools (including MCP tools and API operations), and agents rely on tool recommendation for tool discovery during task execution.
Figure~\ref{fig:topk_recall_vs_traces} shows consistently high recall, Top-10 recall exceeds 96\%, Top-15 and larger recommendation sets exceed 98\%.

\revise{
    The offline rule extraction cost is amortized over the $Q$ queries between two tool-set changes; the per-query cost of the pipeline is
    \begin{equation}
        C_{\text{RE-Agent}} = \frac{C_{\mathrm{extract}}(\Delta N) + C_{\mathrm{deploy}}}{Q}
        + C_{\mathrm{rewrite}} + C_{\mathrm{retrieval}}
    \end{equation} 
    where $C_{\mathrm{extract}}$ grows only with the changed tools $\Delta N$. 
    In our measurements, a full pass over 3{,}616 tools takes ${\sim}$60 s and yields ${\sim}$200 rules.
    The online rewriting adds a median of ${\sim}$560ms.
    The cost is not very high.
    This is because a small ${\sim}$3B activated MoE model suffices, and the KV cache of shared rules can be reused across queries.
    This cost avoids extra LLM tool selection round-trips (Table~\ref{tab:e2e_latency}), which are more expensive.
    Currently, the rewriting is enabled per tenant if their workflows contain ID-only/weak-semantic queries;
    otherwise queries use plain hybrid retrieval.
}

\begin{lessonbox}
\textbf{Lesson 4:} \textit{Under-specified queries need domain knowledge, and that knowledge can be extracted automatically.}
\end{lessonbox}



\begin{reviseblock}
    
\section{Discussion}

\noindent\textbf{Cloud gateway design choices.}
We implemented the protocol conversion using WebAssembly (WASM). 
This isolates the pipeline from the host, ensuring that each stage remains modular for fast-evolving requirements.
We deployed the MCP integration in the cloud rather than alongside the client, so credentials can be managed securely instead of being distributed to clients.
Also, services and the retrieval engine scale independently without the agent being aware of it.

\noindent\textbf{OpenAPI conversion limitations.}
Currently, pagination and polling can be handled by the agent via multiple calls, but callbacks initiated by a backend cannot be received by the gateway. We plan to implement an event-bridging layer on the gateway to address this limitation.

\noindent\textbf{Future work for security bridging.}
Bridging ingress and egress authentication makes the gateway a trust boundary that holds backend credentials on behalf of agents.
Current mitigations include fine-grained access control. 
Our next steps involve isolating credential storage by decoupling traffic forwarding from credential management and supporting credential rotation to secure the system.

\end{reviseblock}



\section{Related Work}
\textbf{L7 Load Balancers and Gateways.}
L7 gateways and load balancers are widely deployed in cloud environments.
Prior systems work has studied L7 load balancing and gateway designs in production settings~\cite{pan2025hermes,wei2024qdsr,saokar2023servicerouter}.
There are also open-source L7 gateways and load balancers (e.g., NGINX and Envoy)~\cite{nginx_latest_docs,envoy_latest_docs,haproxy_latest_docs,apisix_latest_docs}.
L7 load balancers and gateways primarily make routing decisions based on HTTP metadata (e.g., host, path, and headers) and typically treat request bodies as opaque.
As a result, they do not natively support parsing JSON-RPC payloads required by MCP, and they cannot directly differentiate tool-invocation requests or have protocol-level capabilities without non-trivial extensions.

\noindent\textbf{LLM Agent Infrastructures.}
Agent infrastructure work studies efficient execution and serving of LLM-based applications, including serving systems for LLM inference~\cite{zheng2024sglang,kwon2023efficient,qin2025mooncake} and scheduling inference engines for agentic programs~\cite{lin2024parrot,luo2025autellix, tan2025towards, chen2025kairos, bian2025tokencake, laju2026nalar}. 
Orthogonal to our focus on scalable tool access for agents, Cortex~\cite{ruan2026cortex} reduces latency and cost for remote knowledge/data access in LLM agents via semantic-aware caching, whereas our MCP Gateway focuses on scalable tool access, multi-version MCP transport support, and routing across heterogeneous tool servers.
Another line of work focuses on agent architectures, such as memory management~\cite{zhong2024memorybank,packer2023memgpt,xu2025mem}, reasoning frameworks~\cite{yao2023react,wei2022cot,yao2023tot}, and agent runtimes~\cite{mei2024aios}, to improve task performance and robustness.
The MCP Gateway targets the tool-calling infrastructure, providing efficient and scalable tool access.
Other studies improve tool selection and argument generation~\cite{schick2023toolformer,patil2024gorilla,qin2024toolllm} and enhance tool-calling reliability under large tool sets~\cite{gan2025rag}.
Recent MCP-related studies examine protocol adoption, reliability, and security~\cite{guo2025systematic}.

\begin{reviseblock}


\begin{reviseblock}

\begin{table}[t]
\revisecolor\centering\footnotesize
\caption{\revise{Comparison with open-source and commercial gateway systems (October 2026). \cfull~supported, \chalf~partial, \cnone~absent or undocumented.}}
\label{tab:gateway_comparison}
\begin{tabular*}{\linewidth}{@{\extracolsep{\fill}}lccccc@{}}
\toprule
 & API & Transport & Auth & Search & Session \\
\midrule
MCP Gateway & \cfull & \cfull & \cfull & \cfull & \cfull \\
AgentCore~\cite{agentcore_gateway_latest_docs} & \cfull & \cnone & \cfull & \chalf & \chalf \\
Microsoft~\cite{microsoft_mcp_gateway_docs} & \cnone & \cnone & \chalf & \cnone & \chalf \\
agentgateway~\cite{agentgateway_dev_latest_docs} & \cfull & \cfull & \cfull & \cnone & \chalf \\
LiteLLM~\cite{litellm_github} & \cfull & \chalf & \cfull & \chalf & \chalf \\
ToolHive~\cite{toolhive_github} & \cnone & \chalf & \cfull & \cfull & \chalf \\
\bottomrule
\end{tabular*}
\end{table}

\end{reviseblock}


\noindent\textbf{MCP Integration gap analysis.}
As shown in \Cref{tab:gateway_comparison}, parallel to the development and deployment of our MCP Gateway, open-source and commercial systems~\cite{ravitemer_mcphub,samanhappy_mcphub,microsoft_mcp_gateway_docs,agentgateway_dev_latest_docs,litellm_github,toolhive_github,agentcore_gateway_latest_docs} have also adopted MCP integration to standardize agent tool access. 
Our work is beyond merely handling the MCP protocol layer; its value lies in its design and measurements tailored to actual MCP workloads and scenarios.

In terms of capabilities, API conversion and authentication are now common; the key difference lies in the granularity.
The first gap is the transport protocol conversion, particularly enabling interaction between HTTP+SSE and Streamable HTTP. 
agentgateway~\cite{agentgateway_dev_latest_docs} and our system achieve this. 
LiteLLM~\cite{litellm_github} supports this by reconstructing a session for every call, which introduces overhead according to our measurements. 
AgentCore~\cite{agentcore_gateway_latest_docs} and the Microsoft MCP Gateway~\cite{microsoft_mcp_gateway_docs} support single or latest transport. 
The second is session-aware routing. 
To the best of our knowledge, MCP Gateway is the first work to present the design and measurements of session affinity and session consistency while decoupling scaling of the gateway layer and the MCP service layer. 
Furthermore, our system handles SSE bridging across multiple gateway instances and backends as a part of session-aware routing. 
The third is the retrieval engine. 
ToolHive~\cite{toolhive_github} and our system offer hybrid retrieval. 
Our advantage comes from production workloads and the corresponding components, e.g., RE-Agent and Knowledge Cache. 

\end{reviseblock}



\section{Conclusion}
Deploying MCP services at cloud scale is not a drop-in change, and traditional L7 load balancers cannot be directly applied to MCP.
This motivates the MCP Gateway System.
The protocol adaptation module translates across multiple protocols and connects agents to large-scale legacy API servers, addressing tool provider-side integration.
Gateway-side function offloading centralizes fine-grained access control via ingress and egress authentication and reduces the burden on MCP developers.
Tool recommendation provides efficient and scalable tool selection, reducing agent tool selection time and token usage, and addressing the agent-side challenge of accessing massive tools.
Session-aware routing enables replicated MCP backends and elastic gateway scale-out for stateful deployments, while the gateway achieves low processing latency by supporting long-lived sessions.
%
Our results show that the MCP Gateway enables massive tool access and scales agent tool access in production.






\begin{acks}

We thank our shepherd and the anonymous reviewers for their insightful feedback. 
This work was supported by the National Key R\&D Program of China under Grant {2023YFB45024\allowbreak00}; 
the National Natural Science Foundation of China under Grants U24B20153, 62402212, 62272223, and U22A2031; 
the Natural Science Foundation for Young Scientists of Jiangsu Province under Grant BK20241245; 
and the Fundamental and Interdisciplinary Disciplines Breakthrough Plan of the Ministry of Education of China under Grant JYB2025XDXM901.
\end{acks}



\bibliographystyle{ACM-Reference-Format}
\bibliography{reference}

\appendix
\raggedbottom

\renewcommand{\thefigure}{A\arabic{figure}}
\renewcommand{\thetable}{A\arabic{table}}
\setcounter{figure}{0}
\setcounter{table}{0}

\newcommand{\tool}[1]{\texttt{#1}}
\newcommand{\cacheoutput}[1]{%
  \begin{takeawaybox}{0.4pt}{4pt}{4pt}%
    \textbf{Knowledge Cache}\par\smallskip
    {\raggedright\ttfamily\small #1\par}%
  \end{takeawaybox}%
}

\section*{Appendix}

\section{MCP Interaction Workflow} \label{sec:mcp-interaction-workflow}

\begin{figure}[htbp]
  \centering
  \includegraphics[width=0.9\linewidth]{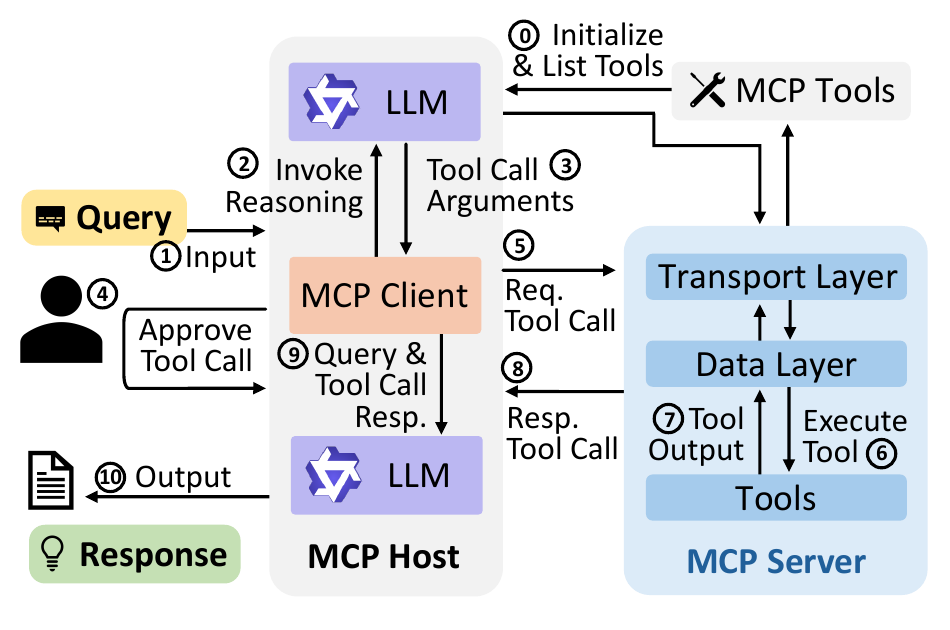}
  \caption{
    MCP interaction workflow
  }
  \Description{The MCP host initializes a client and discovers tools, accepts a user query, invokes the LLM to select tools and generate arguments, obtains approval, calls the MCP server, and returns tool results to the LLM until it can produce the final response.}
  \label{fig:mcp_workflow}
\end{figure}

The MCP host, i.e., an LLM application or agent, follows a one-client-per-server model: it instantiates and maintains a dedicated MCP client instance for each MCP server it connects to.

As illustrated in \Cref{fig:mcp_workflow}, from the tool-calling perspective, the interaction begins with connection establishment and initialization. 
The client performs tool capability discovery by invoking the \textit{List Tools} primitive to obtain the server's complete list of tool definitions (\circled{0}).
Typically, the tool list is cached on the host.
Upon receiving a user query (\circled{1}), the MCP host constructs a composite prompt that includes both the query and the discovered tool specifications.
The specifications for all discovered tools, i.e., their names, descriptions, and parameter schemas, are serialized within the composite prompt.
Leveraging the reasoning capabilities of the LLM, the host invokes the LLM to select the appropriate tools (\circled{2}) and generates structured arguments (\circled{3}) for the selected tools.
The host parses the structured tool invocation arguments returned by the LLM and prepares to initiate the tool call via the client.
To reduce execution risk, the host may request user authorization to execute the target tools (\circled{4}).
Once approved, the client issues a Tool Call primitive as a JSON-RPC request (\circled{5}), triggering local tool execution on the MCP Server (\circled{6}--\circled{7}).
The server returns the tool execution output to the client (\circled{8}), and the host appends it to the ongoing LLM interaction context (\circled{9}).
The LLM then uses the updated context to determine whether the user request has been satisfied; if not, the agent repeats steps \circled{2}--\circled{9} until completion, at which point the LLM generates the final response (\circled{10}).

\section{Handling Long-Lived Connection}
\label{sec:long-lived-connection-handling}

\begin{figure*}[!t]
    \centering
    \includegraphics[width=0.9\linewidth]{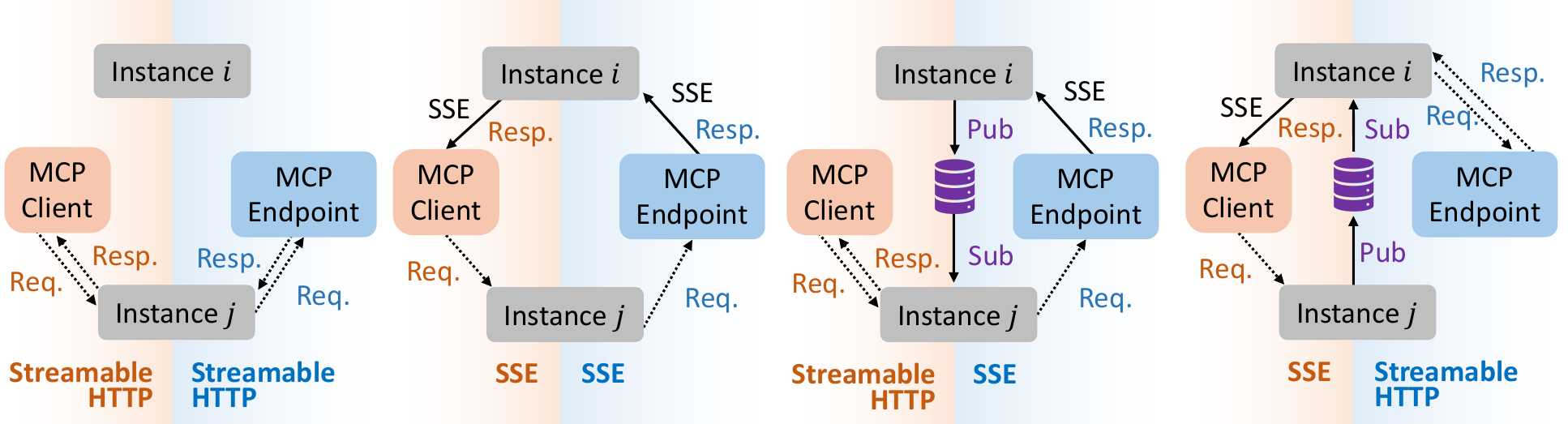}
    \caption{Handling long-lived SSE connections and HTTP response}
    \Description{
        Four panels show frontend/backend transport combinations: Streamable HTTP/Streamable HTTP, HTTP+SSE/HTTP+SSE, Streamable HTTP/HTTP+SSE, and HTTP+SSE/Streamable HTTP. The mixed-transport panels use Pub/Sub to relay responses or requests between gateway instances so that responses reach the instance owning the required return channel.
    }
    \label{fig:sse-long-live}
\end{figure*}

Transport modes impose different constraints on session consistency:
In Streamable HTTP transport, SSE streams are typically request-scoped and closed after the response is delivered, so long-lived channel ownership by a specific gateway instance is not required.
In HTTP+SSE transport, the SSE channel is long-lived and strongly bound: the client maintains a persistent SSE connection to a particular gateway instance, and all server-to-client responses and notifications must traverse that channel.
This introduces a structural hazard: a subsequent request carrying the same Session-ID:Frontend may be load-balanced to a different gateway instance that does not own the SSE return channel.

\balance

To address this, as shown in \Cref{fig:sse-long-live}, the gateway uses a cross-node \textbf{Pub/Sub} coordination mechanism and the Session Metadata Store mapping between Session-ID:Frontend and Session-ID:Backend.

Overall, the gateway makes the relationship between ``request accessing instance'' and ``response channel ownership'' explicit.
Only when a mismatch could break end-to-end delivery does it use Pub/Sub to transfer execution or relay responses across nodes.
This preserves session-centric semantics and availability for long-lived connections while keeping the system elastically scalable.

\section{Prompt and Request Settings for Figure~9}
\label{app:figure9-prompt}

This appendix supplements the experimental protocol of
Figure~\ref{fig:retrieval-comparison}.
All methods use the same system prompt and the original
user query at the LLM tool-selection stage, we omit the sensitive information in the system prompt, and provide an example here, the real prompts are complex than this example, but the structure is similar.:

\begin{quote}
\small
"You are a tool-calling agent ... select the single most
appropriate tool for the user request and call it..."
\end{quote}

The following example illustrates the request configuration
used in the evaluation:

\begin{verbatim}
request = {
    "model": "qwen3.6-plus",
    "temperature": 0,
    "enable_thinking": False,
    "tool_choice": "required",
    "messages": [
        {"role": "system",
         "content": system_prompt},
        {"role": "user", "content": query}
    ],
    "tools": candidate_tools
}
\end{verbatim}

\looseness=-1
The \texttt{candidate\_tools} list contains the full tool pool
for Expose-all and the retrieved Top-15 candidates for
retrieval-based methods.
\clearpage
\nobalance
\onecolumn
\thispagestyle{plain}

\begingroup
\setlength{\parindent}{0pt}
\setlength{\parskip}{0.7\baselineskip}
\pdfbookmark[0]{Statement on Changes and Corrections}{statement-corrections}
\begin{center}
  {\Large\bfseries Statement on Changes and Corrections\par}
  {\normalsize SIGOPS ATC Version\par}
\end{center}
\noindent\rule{\linewidth}{0.4pt}\par

This version adds supplementary appendix material on basic MCP concepts and how we bridge the SSE return path.

Most importantly, it corrects the evaluation results and baseline descriptions in Figure~\ref{fig:retrieval-comparison}.
The Expose-all baseline in Figure~\ref{fig:cmp-tb-e2e}
was inadvertently evaluated under a policy that allowed
responses without tool calls, as in the experiment shown in Figure~\ref{fig:llm_accuracy}.
While this policy allows the model to decide whether to invoke
a tool, our evaluation focuses on selecting the correct tool
for queries that require a tool call.
We correct this mismatch by re-evaluating Expose-all with \texttt{tool\_choice=required}
and the same prompt used for all other methods (see Appendix~\ref{app:figure9-prompt}) to ensure a consistent tool-calling policy across methods.
We also add Appendix~\ref{app:figure9-prompt} to clarify
the experimental details omitted from the ATC version,
including the system prompt, request configuration, and
how candidate tools are supplied to each method.
We update the corresponding baseline descriptions,
results, and discussion.

The most notable change is the increase in the reported
Expose-all selection accuracy (Figure~\ref{fig:cmp-tb-e2e}).
This correction does not change the paper's main conclusions.
The Expose-all baseline remains inaccurate at large scale (Figure~\ref{fig:cmp-ali-e2e}).
The MCP Gateway substantially reduces tool-selection latency
and token usage while maintaining high retrieval recall
and tool-selection accuracy on production workloads.
The reported latency and token-usage measurements remain unchange.

We thank the readers for their careful feedback, which helped make this paper more accurate.
\par
\endgroup

\end{document}